\documentclass[pra,a4paper,superscriptaddress,twocolumn,amsmath,amssymb,floatfix,amstex]{revtex4}%

\usepackage{graphicx}
\usepackage{grffile}
\usepackage{dcolumn}
\usepackage[hidelinks]{hyperref}
\usepackage{mathptmx}
\usepackage{verbatim}
\usepackage{mathrsfs}
\usepackage{xcolor}
\usepackage{physics}
\usepackage{ulem}
\usepackage[T1]{fontenc}
\usepackage[utf8]{inputenc} 
\usepackage[english]{babel}    % pour un style de mise en page francais et anglais
\usepackage{fixltx2e} %sinon les figures sur une seule colonne sont mal numéheolootées
\usepackage{mathbbol} %pour avoir \mathbb{1} correct
\usepackage{soul}

\def\cg{\mathcal{G}}
\def\cp{\mathcal{P}}
\def\calc{\mathcal{C}}

\usepackage{listings}

\newcommand{\SL}[1]{{\color{black} #1}}
\newcommand{\SLNB}[1]{{\color{black} #1}}

\renewcommand{\leq}{\leqslant}

\begin{document}

\title{%Network models for the cerebral vasculature 
%Community structure and network resilience to localized damage: \\ application to brain microcirculation

%Community structure and resilience to localized damage: \\ application to brain microvascular networks

Network community structure and resilience to localized damage: \\ application to brain microcirculation}

\author{F.~Goirand}
\affiliation{Institut de Mécanique des Fluides de Toulouse (IMFT), Université de Toulouse, CNRS, Toulouse, France}
\affiliation{Univ Rennes, CNRS, G\'eosciences Rennes, UMR 6118, 35000 Rennes, France.}
\author{B.~Georgeot}
\affiliation{Laboratoire de Physique Th\'eorique, Universit\'e de Toulouse, CNRS, UPS, France}
\author{O.~Giraud}
\affiliation{Universit\'e Paris Saclay, CNRS, LPTMS, 91405, Orsay, France}
\author{S.~Lorthois}
\affiliation{Institut de Mécanique des Fluides de Toulouse (IMFT), Université de Toulouse, CNRS, Toulouse, France}

\begin{abstract}
%\SLNB{SL: A relire à la fin}

%We study the topological structure of cerebrovascular networks, using a biological network from a mouse brain as testbed, and modelize them using a variant of random regular graphs modified to include substructures (communities). We show that such models can describe the main topological features of the biological network, and enable to probe their properties. In particular, our results indicate that the distribution of flow reductions when an edge is removed (i.e.~a vessel is obstructed  in a biological setting) is strongly influenced by the community structure: when the community structure gets stronger, the probability of large flow reductions increases substantially, with potential biological consequences.

In cerebrovascular networks, some vertices are more connected to each other than with the rest of the vasculature, defining a community structure. Here, we introduce a class of model networks built by rewiring Random Regular Graphs, which enables \SLNB{reproduction of} %to reproduce 
this community structure and other topological properties of cerebrovascular networks. We use these model networks to study the global flow reduction induced by the removal of a single edge. We analytically show that this global flow reduction can be expressed as a function of  the  initial  flow  rate  in  the  removed edge and of a topological quantity, both of which display probability distributions following Cauchy laws, i.e.~with large tails.   As a result,  we show that the distribution of blood flow reductions is strongly influenced by the community structure. In particular, the probability of large flow reductions increases substantially when the community structure is stronger, weakening the network resilience to single capillary occlusions. We discuss the implications of these findings in the context of Alzheimer’s Disease, in which the importance of vascular mechanisms, including capillary occlusions, is beginning to be uncovered.
\end{abstract}
\date{\today}

\maketitle

%%%%%%%%%%%%%%%%%%%%%%%%%%%%%%%%%%%%%%%%%%%%%%%%%%%%%%%%%%%%%%%%%%%%%%%%%
%%%%%%%%%%%%%%%%%%%%%%%%%%%%%%%%%%%%%%%%%%%%%%%%%%%%%%%%%%%%%%%%%%%%%%%%%
\section{Introduction}\label{introduction}
%%%%%%%%%%%%%%%%%%%%%%%%%%%%%%%%%%%%%%%%%%%%%%%%%%%%%%%%%%%%%%%%%%%%%%%%%
%%%%%%%%%%%%%%%%%%%%%%%%%%%%%%%%%%%%%%%%%%%%%%%%%%%%%%%%%%%%%%%%%%%%%%%%%

Cerebral hypoperfusion, i.e.~the decrease of cerebral blood flow, is a common feature of many brain diseases, including neurodegenerative diseases, such as Alzheimer's Disease (AD)~\cite{Attwell_review2020,Cortes2020}, and cerebrovascular diseases, such as hypoperfusion dementia~\cite{Iadecola2013}. %\textcolor{blue}{(add ref ?)}
Hypoperfusion %, whether localized or systemic (better word ?),
 is a key player in the onset and progression of cerebrovascular diseases~\cite{Iadecola2013, Dong2018} and has been considered until recently as a consequence of %synapse and neuron loss induced by the accumulation of neurotoxic metabolic waste in AD brains, including amyloid $\beta$ and tau proteins
neurodegeneration in AD~\cite{Iadecola2010}. However, this view is now debated~\cite{Attwell_review2020}. In human patients, cerebral blood flow indeed decreases — in a statistical and epidemiologic sense — before neurotoxic waste accumulate in the brain and before any measurable cognitive deficits~\cite{Iturria}. Moreover, occlusions of capillary vessels by white blood cells (neutrophils) have been observed in animal models of AD before the accumulation of amyloid~$\beta$, the main neurotoxic protein forming deposits (plaques) in AD brains~\cite{cruz2019neutrophil}. Despite the small proportion of occluded vessels (\SLNB{from} 1 to 4\%), the pharmacological removal of these neutrophils led to a significant increase in blood flow and improved the cognitive performance of the animals. At a later stage , i.e.~when the animals already showed plaques, extensive capillary constrictions have also been observed~\cite{NortleyScience}. The \SLNB{exposure} of pericytes, i.e.~active mural cells wrapping around the capillaries, to increased concentrations of amyloid~$\beta$ has been shown to induce their contraction. This contributes to a positive feedback %\textcolor{blue}{or self-reinforcing}
loop, where decreased cerebral blood flow not only triggers biological pathways leading to increased amyloid~$\beta$ production in the brain, but also directly impairs its elimination by the flowing blood. This results in increased amyloid~$\beta$ accumulation in the brain, increased pericyte contraction and further hypoperfusion~\cite{NortleyScience}. In parallel, hypoperfusion also directly compromises the brain's energy supply, with deleterious neurological consequences.

A central question in this context is: to what extent could a small proportion of vessel occlusions trigger the above positive feedback loop, contributing to AD onset and progression? Using highly resolved simulations of blood flow in anatomically realistic microvascular networks from human and mice, it was previously shown that, on average, cerebral blood flow decreases linearly with an increasing proportion of capillaries occluded at random, up to 20\%, i.e.~without any threshold effect~\cite{cruz2019neutrophil}. Thus, on average, each single capillary occlusion has a similar, and cumulative, contribution to the blood flow decrease at the scale of the network. Here, we focus on the variability of this contribution %. For that purpose, we take a step back and adopt a theoretical point of view, where we
 and seek to determine how it is controlled by the fundamental topological properties of cerebrovascular networks. In particular, we investigate the role of network communities, i.e.~substructures with vertices more connected to each other than to other vertices, which have been identified in such networks~\cite{Blinder2013,glioblastoma} and many other real-life networks \cite{ForCas,fortunato2010}. %Communities are an ubiquitous feature of many real-life networks, and correspond to groups of vertices that are more likely to be connected together than with the rest of the graph.
In social networks, for example, these communities represent groups of users with specific affinities, the identification of which is an important area of research~\cite{social1, social2}. Such communities are also important to understand the organization of the World Wide Web~\cite{web1, web2}, as they influence the various page ranking algorithms.

 %\SLNB{ (OG or BG : could you also re-include citations that arrive later in the text and may be relevant for communities in other real-life networks ? e.g.~from~\cite{AlbertBarabasi, boccaletti,dorogovtsev,bollobas, albert, dorog2003b} ?) }%\OB{However, many real-life networks, and the cerebrovascular networks~\cite{Blinder2013,glioblastoma} in particular, include substructures, or communities, groups of vertices more connected to each other than to other vertices.}
%. For that purpose, we take a step back and adopt a theoretical point of view, where we consider the occlusion of a single vessel in model networks of increasing complexity, but where all vessels have identical unit conductivity. In other words, by contrast to~\cite{Hudetz1993,Pozrikidis2012}, we ignore the contribution of vessel morphology, including distributions of diameters and lengths. We also ignore the complex rheology of blood. 
In order to study the role of communities in brain microvascular networks, we take a step back and adopt a theoretical point of view, using model networks of increasing complexity, as illustrated in Fig.~\ref{examplesgraphs}, but with 
a single inlet and outlet, and 
vessels with identical unit conductivity. On the one hand, this  abstract approach enables \SLNB{the network community structure to be controlled}. On the other hand, it enables \SLNB{analytical expressions to be derived}, or \SLNB{averages over many realizations of similar graphs with same properties to be performed}, in order to analyze the variability of the blood flow reduction induced by the occlusion of a single vessel. %On the one hand, this  abstract approach enables to control the network community structure. On the other hand, it enables to derive analytical expressions, or to perform averages over many realizations of similar graphs with same properties, in order to analyze the variability of the blood flow reduction induced by the occlusion of a single vessel. %To investigate these questions, we will use different specific network models, which enable to probe the consequence of the topology and structures on the blood flow. Here we will base our investigations on the following ones. First, 
We first consider simple ideal graphs known as Random Regular Graphs (RRGs) in network theory~\cite{wormald99}. These graphs do  not  need  to  be  embedded  in  the  physical  space and they are structureless. They  nevertheless reproduce one of the main topological properties of cerebrovascular networks, in which most vertices have the same \SLNB{connectivity} (or degree), equal to three (see e.g.~\cite{Blinder2013,Tsai2009,Smith2019}). Moreover, many realizations of RRGs of arbitrary size can be easily constructed, and they locally behave like trees, enabling analytical derivations that provide insight on their asymptotic behavior in the limit of large sizes. % \OB{However, many real-life networks, and the cerebrovascular networks~\cite{Blinder2013,glioblastoma} in particular, include substructures, or communities, groups of vertices more connected to each other than to other vertices. To investigate their effect, }
We then modify this ideal RRG model to  %account for such community structure and
provide a simple generation scheme that enables \SLNB{ the strength of the communities to be controlled} by rewiring together a finite number of elementary RRGs. As a third model, we use random networks constructed from Voronoi diagrams of sets of points homogeneously distributed in 3D space, following~\cite{Smith2019}. Such spatial networks are locally randomized but homogeneous at the network scale, and reproduce both the structure \SL{(morphology and topology)} and function \SL{(flow, blood/tissue exchange and
robustness to capillary occlusions)} of brain capillary networks. Finally, we also consider the intracortical vascular network from the mouse parietal cortex (15,000 vessels in a 1$mm^3$ region) used in~\cite{cruz2019neutrophil}. In these last two anatomically realistic networks, by contrast with~\cite{Hudetz1993,Pozrikidis2012}, we neglect the contribution of vessel morphology, including distributions of diameters and lengths, and impose unit conductivity in all edges.

We use the above models to %investigate the fluxes across the network, and their relation to topology. In particular, we
study the impact of single edge removal, i.e.~equivalent to vessel occlusion. %obstruction.
 We show that the resulting flow reduction at network scale can be expressed as a function of the initial flow rate in the occluded %obstructed
 vessel and a topological quantity, both of which display probability distributions with large tails. As a result, we show that the distribution of blood flow reductions may display unexpectedly large values, the probability of which increases when the community structure is stronger. Such results indicate that the topology of biological
networks, including their community structure, is important to assess their functional properties, especially their biological
resilience \SL{understood here as their ability to maintain functionality in the event of localized damage~\cite{gavrilchenko_resilience_2019}}.
%SL : améliorer ? is important to ? flow reduction vs flow reduction

The paper is organized as follows. In Section II, we introduce the network models that will be considered, and investigate in
Section III their topological properties compared to  anatomically realistic
networks. In Section IV, we consider %the properties of 
blood flow %fluxes 
through these networks. We uncover in particular the quantities controlling
the distribution of flow %flux 
reductions induced by the removal of a single edge %edges are 
 and highlight how the network topological properties influence this distribution. %In Section IV, we consider the properties of fluxes through the network, highlighting in particular the distribution of flow reductions when edges are removed, and the importance of the topological properties for such quantities.  
 Finally, in Section V, we discuss these findings and their implication for brain pathophysiology.
 
%We then present some conclusions and links to the biological consequences of our findings. \SLNB{(A revoir)} 

\bigskip
\section{Network models}\label{definitions}
%%%%%%%%%%%%%%%%%%%%%%%%%%%%%%%%%%%%%%%%%%%%%%%%%%%%%%%%%%%%%%%%%%%%%%%%%
%%%%%%%%%%%%%%%%%%%%%%%%%%%%%%%%%%%%%%%%%%%%%%%%%%%%%%%%%%%%%%%%%%%%%%%%%
Network models have been used in many fields~\cite{AlbertBarabasi, boccaletti,dorogovtsev,bollobas, albert, dorog2003b}, where they proved useful to distinguish the main architectural properties of real-life complex networks that strongly impact their function from peculiarities which may depend on particular instances but do not change much the function.
%Real-world cerbrovascular networks are hard to map experimentally and only partial data can be accessed with great efforts. It is therefore important to devise and build models of such networks, which will keep the main features of the biological ones while enabling more thorough analytical and numerical studies. Another important point of the construction of such models is to distinguish in real networks the important properties which have biological consequences to peculiarities which may depend on particular instances or species and do not change much the properties.

%\OB{To study the microvascular network, we will use network models. This tool has been successfully applied in many fields~\cite{AlbertBarabasi, boccaletti,dorogovtsev,bollobas, albert, dorog2003b}.}
Here, we start from the observation that, as stated in the Introduction, %one of the main topological properties of cerebrovascular networks, in which most vertices have the same coordination number (or degree), equal to three (see e.g.~
% microvascular networks within the brain cortex have the property that 
most vertices in cerebrovascular networks have three neighbours (either one vessel branches into two other vessels or two vessels merge into another one)~\cite{Blinder2013,Smith2019}. The simplest graph model one can think of to study such networks is a graph model only reproducing this feature, i.e.~with constant connectivity. Such graphs are known as regular graphs. Thus, the first model we consider is {\it random regular graphs} (RRGs), where each vertex has the same connectivity $z$, as illustrated in Fig.~\ref{examplesgraphs}, left (see Appendix~\ref{construction} for details). 

RRG graphs by construction do not present specific substructures. However, in many real-life networks, including cerebrovascular networks, groups of vertices may have more links to each other than to other vertices~\cite{Blinder2013,glioblastoma}, defining substructures, or communities. %Such communities are very important e.g.~in social networks or the World Wide Web.
In order to account for such communities, we also consider a slightly more elaborate model, which we call {\it rewired RRG} model. It is obtained by generating $n_c$ independent RRG graphs, which may all have the same size or have a given heterogeneous size distribution, and rewiring pairs of edges at random, as illustrated in Fig.~\ref{examplesgraphs}, middle (see Appendix~\ref{construction} for details). 
In particular, we will denote by $\mathcal{R}_k$ a subfamily of rewired RRGs built from a set of RRGs whose size is distributed to reproduce the communities of cerebrovascular networks, as further introduced in Section \ref{seccomm}, and with $k$ rewirings.

Both RRG and rewired RRG models are ideal graphs of infinite dimension~\cite{AlbertBarabasi,wormald99,kim2003,boccaletti}, which are not embedded in the physical space (i.e.~vertices have no a priori physical spatial coordinates). Thus, to account for the 3D structure of the brain, we consider a phenomenological model, constructed from edges of Voronoi cells in the three-dimensional space, which reproduce both the structure and function of brain capillary networks~\cite{Smith2019} (see Appendix~\ref{construction} for details on the construction). The topology of a typical realization $\mathcal{V}$ of such graphs, that we call {\it Voronoi graphs}, is displayed in Fig.~\ref{examplesgraphs}, right. 

Finally, we also construct a graph based on experimental data from a mouse brain previously acquired by~\cite{Tsai2009,Blinder2013} (see Appendix ~\ref{construction} for details). By contrast to the previous graph, this experimental graph not only includes the space-filling capillary vessels, but also the tree-like penetrating arterioles and ascending venules~\cite{Nguyen2011}. Noteworthy, because these latter graphs are obtained from real 3D structures, vessels intersecting the edges of the considered region are cut. We recover the 3-connectivity of the graph by removing recursively all dangling vertices (vertices of  connectivity one), and we keep only the largest connected component.

\begin{figure}[t!]
\begin{center}
\includegraphics*[width=0.34\linewidth]{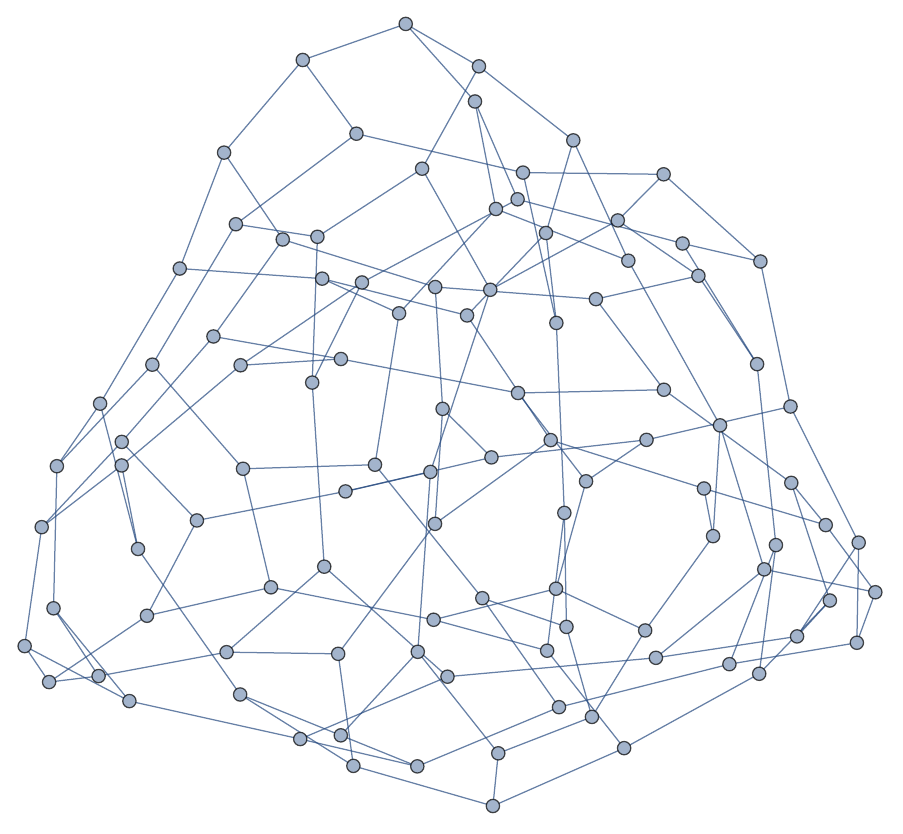}\hfill \includegraphics*[width=0.22\linewidth]{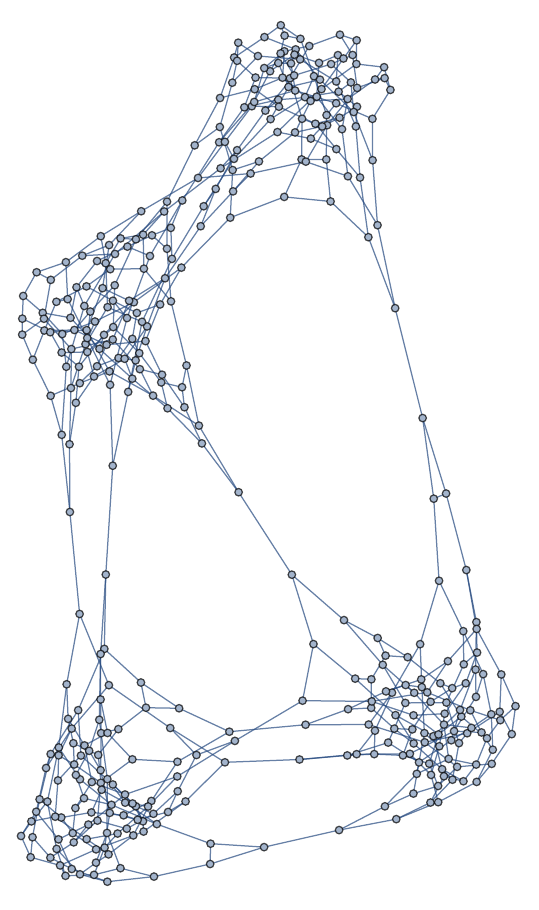}\hfill\includegraphics*[width=0.41\linewidth]{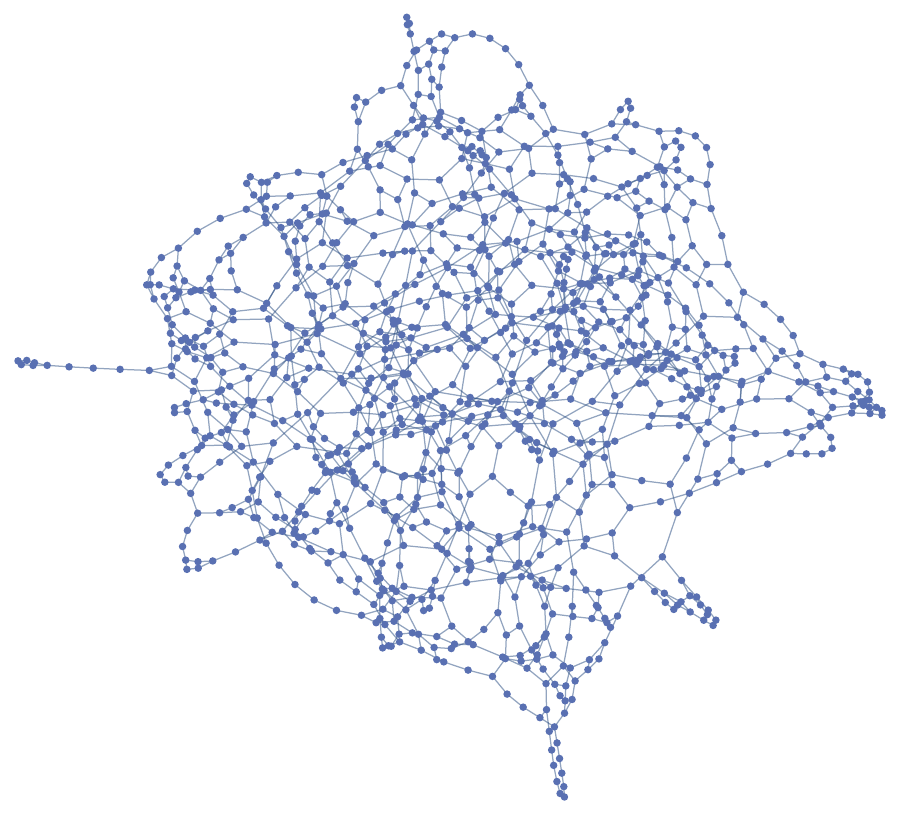}
\caption{\textbf{Topological representation of typical realizations of various graph models}. From left to right: RRG with connectivity $z=3$ and $N=100$ vertices; rewired RRG obtained with 
$n_r=10$ rewirings, starting from 4 RRGs of size $n_0=100$; Voronoi graph $\mathcal{V}$ with $N=1158$ vertices. Note that the figure only represents the existing connections (edges) between vertices, without accounting for their spatial position.}
\label{examplesgraphs}
\end{center}
\end{figure}

%%%%%%%%%%%%%%%%%%%%%%%%%%%%%%%%%%%%%%%%%%%%%%%%%%%%%%%%%%%%%%%%%%%%%%%%%
%%%%%%%%%%%%%%%%%%%%%%%%%%%%%%%%%%%%%%%%%%%%%%%%%%%%%%%%%%%%%%%%%%%%%%%%%
\section{Topological properties}\label{topology}
%%%%%%%%%%%%%%%%%%%%%%%%%%%%%%%%%%%%%%%%%%%%%%%%%%%%%%%%%%%%%%%%%%%%%%%%%
%%%%%%%%%%%%%%%%%%%%%%%%%%%%%%%%%%%%%%%%%%%%%%%%%%%%%%%%%%%%%%%%%%%%%%%%%

%%%%%%%%%%%%%%%%%%%%%%%%%%%%%%%%%%%%%%%%%%%%%%%%%%%%%%%%%%%%%%%%%%%%%%%%%
\subsection{Distribution of loop lengths}
%%%%%%%%%%%%%%%%%%%%%%%%%%%%%%%%%%%%%%%%%%%%%%%%%%%%%%%%%%%%%%%%%%%%%%%%%

\begin{figure}[!h]
\begin{center}
\includegraphics*[width=0.99\linewidth]{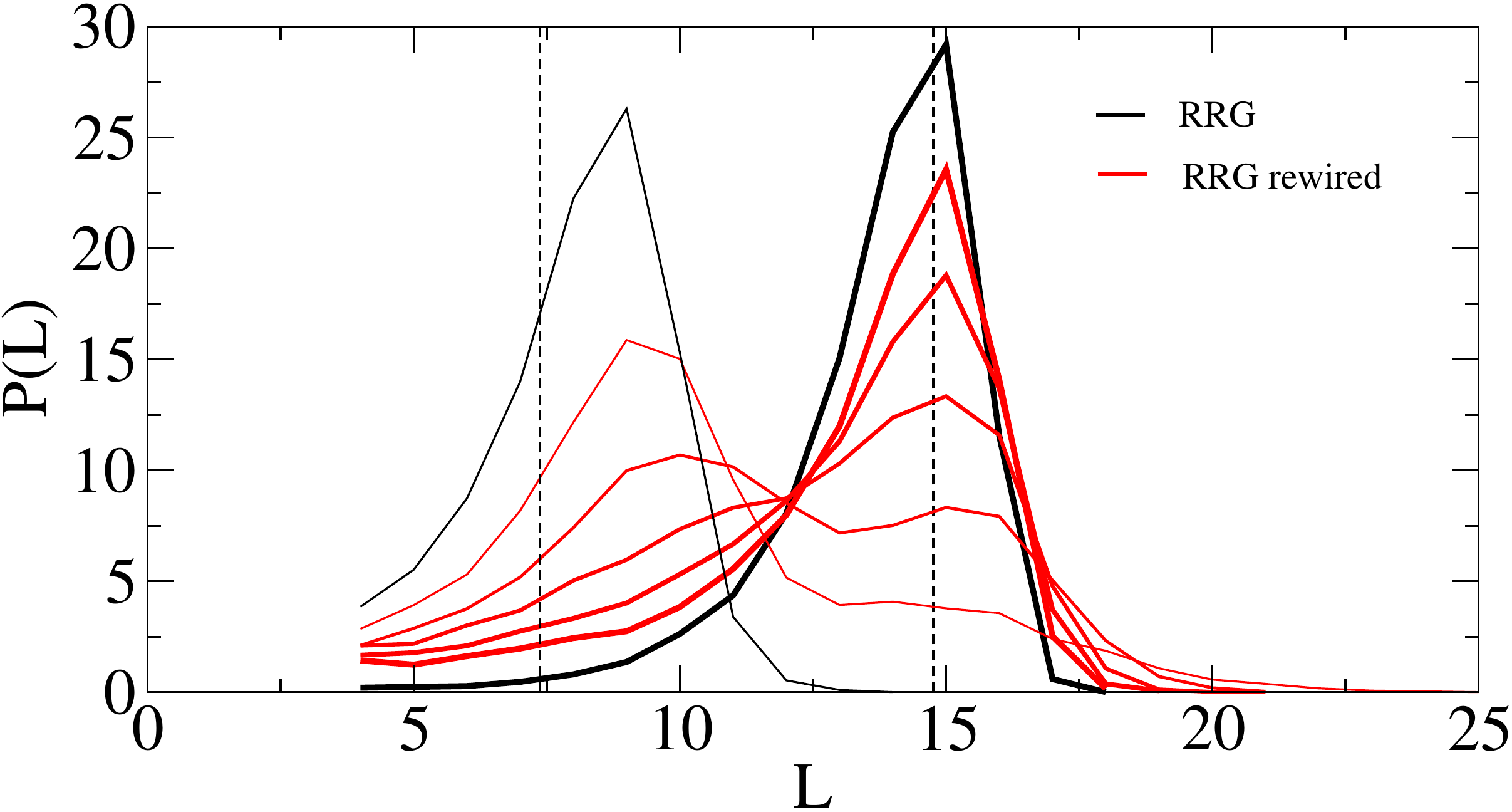}\\
\includegraphics*[width=0.99\linewidth]{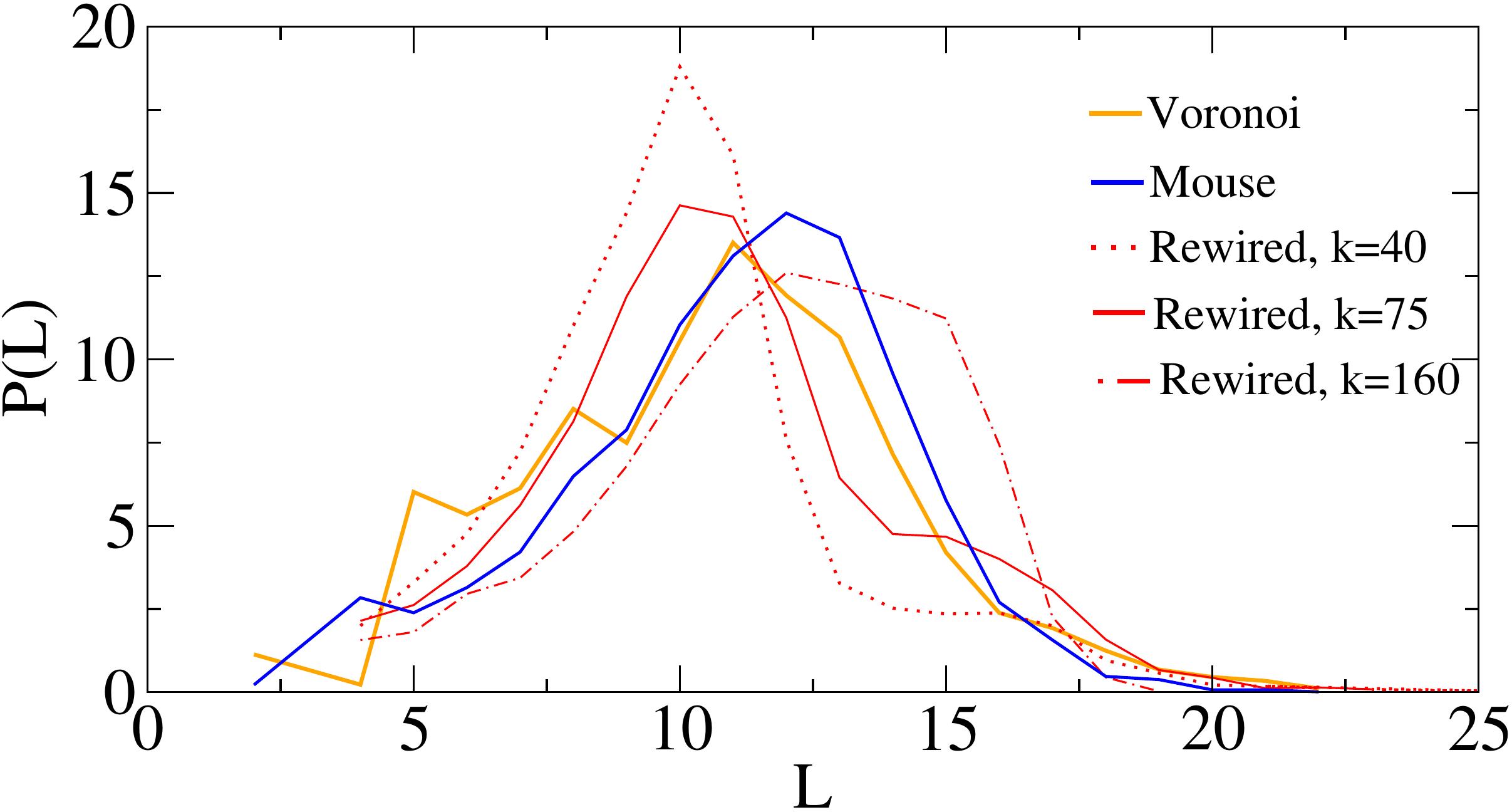}
\caption{\textbf{Distribution of loop lengths.} \textbf{Top: Average loop length distribution for RRGs (black) with $N=1600$ (right, \SL{thick}) or $N=40$ (left, \SL{thin}), and rewired RRGs with 40 elementary RRGs of size 40 (red)} and (from \SL{left (thin) to right (thick)}) 100 to 500 rewirings; each histogram is obtained from 10 random realizations. The vertical dashed lines indicate $2\ln 40\simeq 7.38$ and $2\ln 1600\simeq 14.76$. \textbf{Bottom: Distribution of loop lengths for a Voronoi graph and the mouse network, compared with distributions obtained for rewired RRGs $\mathcal{R}_k$ of heterogeneous substructure size.} Orange: Voronoi graph $\mathcal{V}$ of Fig.~\ref{examplesgraphs} left; Blue: Mouse graph; Red: Average for 10 realizations of rewired RRGs $\mathcal{R}_k$, with $k=40$ (dotted), 75 (solid) and 160 (dashed-dotted) rewirings.
\label{looplengths}}
\end{center}
\end{figure}

The topology of a graph 
\SL{is often characterized by its mean connectivity \cite{boccaletti}. However, graphs with mean identical  connectivity may have very different topologies, from binary trees to periodic regular lattices or scale-free graphs. In the latter,} hubs %that 
allow to connect any two vertices with a very short path~\cite{AlbertBarabasi}\SL{, while in trees, %two vertices are usually connected by very long paths/
there are no shortcuts.} %\textbf{que peut-on dire ici , mot redundancy?}.
\SL{The distribution of loop length enables these different behaviors to be distinguished, while getting additional insight on network redundancy. Moreover, it has been carefully studied in brain microvascular networks~\cite{Blinder2013,Smith2019}.}  
%can be characterized by the lengths of the shortest paths connecting pairs of vertices. For instance, in scale-free graphs, %there are 
%hubs %that 
%allow to connect any two vertices with a very short path~\cite{AlbertBarabasi}.

Here, \SL{following~\cite{Smith2019},} we define a loop as the shortest path going from one vertex $i$ to itself through two given neighbours of that vertex. For a vertex with $z$ neighbours, there are $\binom{z}{2}$ loops. The loop length $L$ is given by the total number of vertices (or edges) in the loop.

In Fig.~\ref{looplengths}, we display the loop length distribution $P(L)$ for the considered network models. As shown in the top panel, for a RRG of size $N$, the distribution is peaked around $2\ln N$. \SL{This result is consistent with the RRGs being locally tree-like~\cite{treelike,dorog2003}}, \SL{since in the limit $N\to\infty$, the relative number of loops of fixed size $p$ goes to 0 for any arbitrary integer $p$, making the graph effectively loopless.} In the same way, the loop distribution of a collection of several independent elementary RRGs of size $n_0$ is peaked at $2\ln n_0$. %Consider now the case of rewired RRGs obtained from elementary RRGs which all have the same size $n_0$ (i.e.~graphs with substructures); in the absence of rewiring, the RRGs of size $n_0$ are independent and from the above argument, their distribution is peaked at $2\ln n_0$. 
Under increased rewiring, the distribution progressively shifts towards the one of a RRG of size $N=n_c n_0$. Thus, rewired RRGs continuously interpolate between a distribution of loops peaked at a value only depending on the substructure size $n_0$ to one peaked at a value only depending on the total network size $N$.

Interpreting the loop length distribution of the Voronoi and mouse networks is less trivial.  As shown in the bottom panel of Fig.~\ref{looplengths}, orange and blue lines, respectively, they are peaked at a similar value, irrespective of their different network sizes, suggesting that, for these networks, the dominant factor is the community size and not the network size. Moreover, the mouse network has a slightly larger proportion of loops with lengths above the peak compared to the Voronoi graph $\mathcal{V}$ (orange line). This is consistent with earlier findings that the Voronoi graphs reproduce well the loop length distribution of cerebral capillary networks~\cite{Smith2019} and the additional contribution of tree-like arterioles and venules in the mouse network~\cite{Lorthois_JTB_2010,Blinder2013}. However, these distributions are larger than the typical distributions obtained in Fig.~\ref{looplengths} top with rewired RRGs of equal size. Such distributions can be roughly reproduced by considering elementary RRGs with heterogeneous sizes (red lines in bottom panel). 
%As shown in the bottom panel of Fig.~\ref{looplengths}, the mouse network (blue line) has a slightly larger proportion of loops with lengths above the peak compared to the Voronoi graph $\mathcal{V}$ displayed in Fig.~\ref{examplesgraphs} right (orange line). This is consistent with the previous finding that the Voronoi graphs reproduce well the loop length distribution of cerebral capillary networks~\cite{Smith2019}, and the additional contribution of tree-like arterioles and venules in the mouse network~\cite{Lorthois_JTB_2010}. 
%However, these distributions are larger than the typical distributions obtained with rewired RRGs of equal size. Such distributions can be roughly reproduced by considering elementary RRGs with heterogeneous sizes (red lines in bottom panel). 
%The Voronoi and mouse networks are peaked at a similar value, irrespective of their different network sizes, suggesting that indeed the dominant factor is the community size and not the network size. 
%Additionally, the results shown in  Fig.~\ref{looplengths}.
This suggests that the simple proposed model of RRG rewiring is sufficiently versatile to generate ideal graphs reproducing the topological properties of intracortical microvascular networks, by contrast to single RRGs, which do not account for the underlying substructures. In the next  subsection, we thus focus on the community structure of these different networks.

%%%%%%%%%%%%%%%%%%%%%%%%%%%%%%%%%%%%%%%%%%%%%%%%%%%%%%%%%%%%%%%%%%%%%%%%%
\subsection{Communities}\label{seccomm}
%%%%%%%%%%%%%%%%%%%%%%%%%%%%%%%%%%%%%%%%%%%%%%%%%%%%%%%%%%%%%%%%%%%%%%%%%

\begin{figure}[!t]
\begin{center}
\includegraphics*[width=0.5\linewidth]{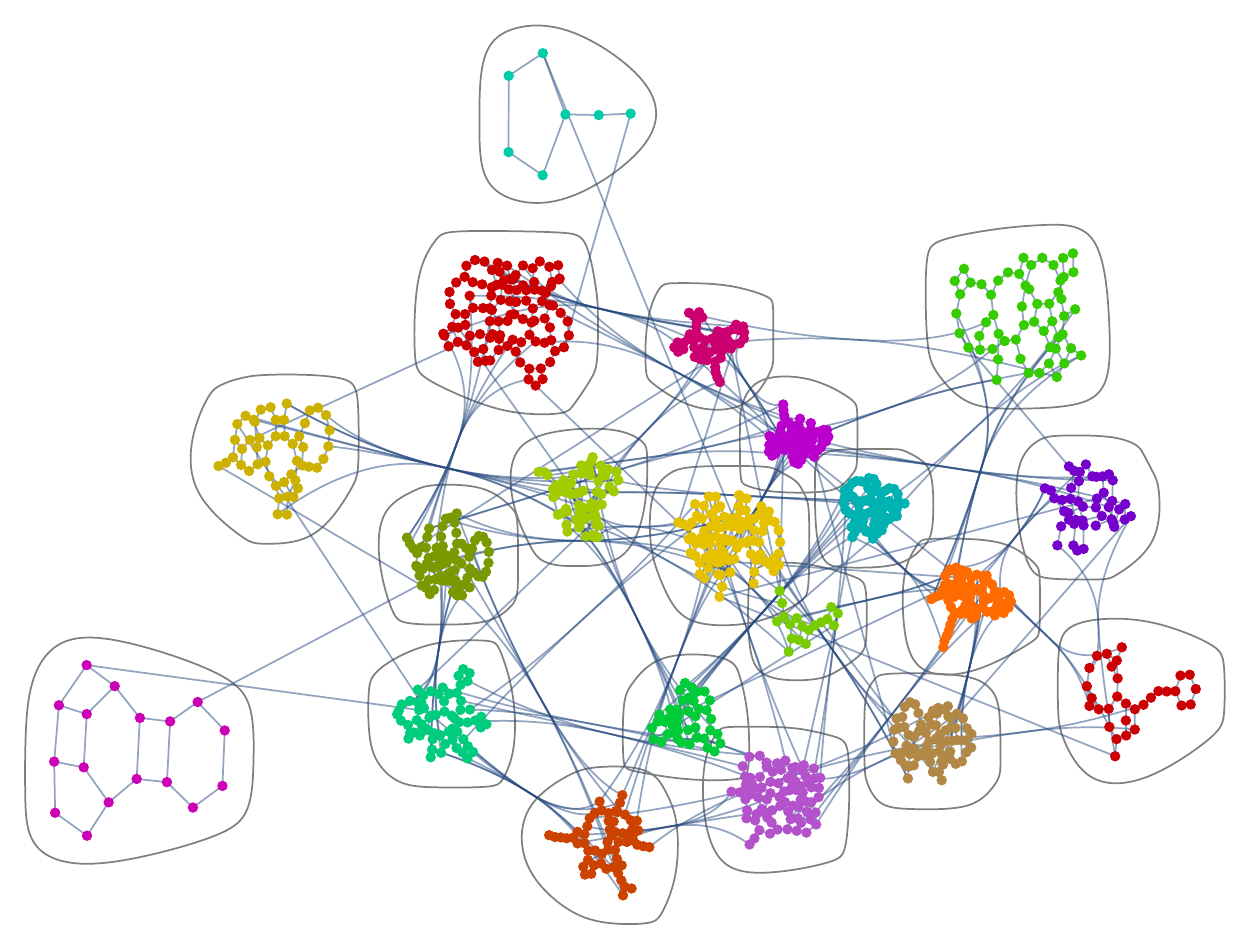}\hfill
\includegraphics*[width=0.5\linewidth]{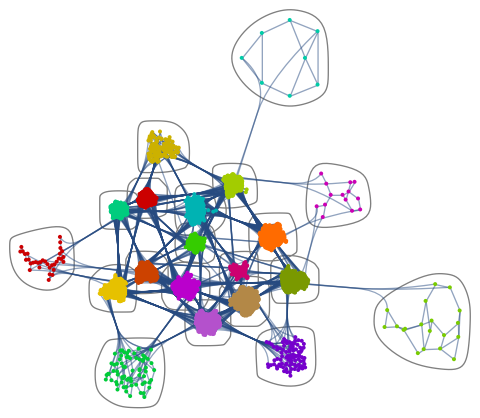}
\caption{\textbf{Community graph plot of the Voronoi graph $\mathcal{V}$ (left) and the mouse graph (right).} The 1158 vertices of $\mathcal{V}$ distribute into 20 communities of sizes 103, 97, 81, 77, 74, 74, 72, 69, 68, 65, 60, 56, 53, 49, 44, 43, 31, 18, 17, 7. The 8720 vertices of the mouse graph distribute into 20 communities of sizes 833, 829, 791, 767, 757, 736, 727, 704, 696, 673, 491, 175, 155, 149, 93, 69, 34, 19, 14, 8.
\label{communautes}}
\end{center}
\end{figure}

%Here, we first examine the community structure of the Voronoi and mouse networks, and %will 
%subsequently discuss how incorporating the findings into the rewired RRG model. 

As mentioned in the Introduction, communities are an ubiquitous feature of many real-life networks, and correspond to groups of vertices that are more likely to be connected together than with the rest of the graph.
%In social networks, for example, these communities represent groups of users with specific affinities, the identification of which is an important area of research~\cite{social1, social2}. Such communities are also important to understand the organization of the World Wide Web~\cite{web1, web2}, as they influence the various page ranking algorithms. 
Such communities have been previously identified in brain microvascular networks~\cite{Blinder2013,glioblastoma}.
To identify these communities%and assess their importance
, we maximize the modularity $\mu (\mathcal{C})$ of all possible partitions $\mathcal{C}$, following~\cite{ForCas,NewmanGirvan04,fortunato2010} and, in the case of intracortical networks,~\cite{Blinder2013,glioblastoma}. The quantity $\mu(\mathcal{C})$ compares the probability of having an edge within a given subset of this partition with the probability expected by chance, i.e.~from connections randomly chosen under the constraint of maintaining the graph connectivity $z$ (see Appendix~\ref{appcommunities} for details). This quantity lies between $-\frac12$ and $1$, and is positive if the partition $\mathcal{C}$ has some relevance as a graph community structure. %should be positive for the community structure to have any relevance. 
We call modularity $\mu$ of the graph the maximal modularity over all partitions, or equivalently the modularity of the optimal partition (see Appendix~\ref{appcommunities}). Noteworthily, even for graphs such as single RRGs without any built-in substructures, the optimization process finds the specific partition of the graph which maximizes the modularity. The value of $\mu(\mathcal{C})$ for this particular partition is usually well above zero. Thus, for single RRGs, we obtain an average modularity $\mu_\textrm{RRG}\sim 0.661$ when averaged over many realizations. This value corresponds to purely random fluctuations, which create random clusters of vertices more tightly bound together than with the rest of the graph. Higher values are expected for the graph community structure to have any relevance. In this case, we directly use the modularity to assess the strength of the obtained community structure. 
For the Voronoi graph we find $\mu_\mathcal{V}\sim$ 0.827 and for the mouse graph $\mu_\textrm{mouse}\sim 0.846$.
Besides modularity, alternative quantities %\OB{besides modularity}
have also been used for that purpose, for example the exponent of the scaling law relating the number of inter- versus intra-community edges% has also been used for that purpose
~\cite{Blinder2013}. However, the range of exponents characterizing strong communities depends on the dimension of the space in which the graph is embedded~\footnote{Following ~\cite{Blinder2013}, the upper-limit corresponds to homogeneous space-filling networks. For such networks in 3D, the number of edges in cubic regions with side $L$ scales as $L^3$, while the number of edges connecting it region to the rest of the network scales as $L^2$. As a result, a power-law with exponent $2/3$ is  obtained for the number of inter-versus intra-community edges. This exponent is $1/2$ in 2D and 1 in infinite dimension.}, making it difficult to compare real-life networks and ideal graphs such as RRGs or rewired RRGs.

%Previous authors have quantified the strength of communities using the power-law behavior of the number of internal links... We prefer... Algorithms [Refs]...

\begin{figure}[!t]
\begin{center}
\includegraphics*[width=0.5\linewidth]{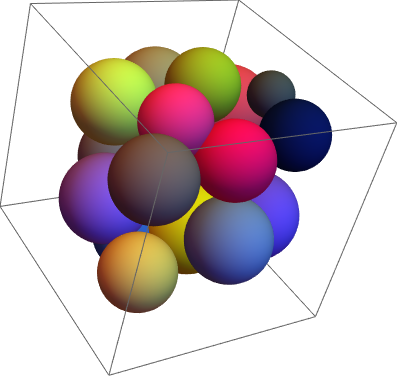}\hfill
\includegraphics*[width=0.5\linewidth]{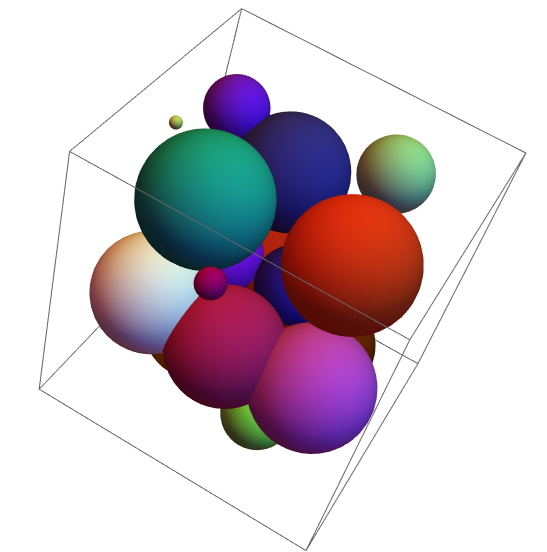}
\caption{\textbf{Spatial localization of communities extracted from the Voronoi graph $\mathcal{V}$ (left) and the mouse graph (right).} Spheres are centered at the barycenter of vertices in a given community, and their radius is given by the standard deviation of vertex positions around that barycenter.
\label{spheresCommunitiessouris}}
\end{center}
\end{figure}

\begin{figure}[!t]
\begin{center}
\includegraphics*[width=0.5\linewidth]{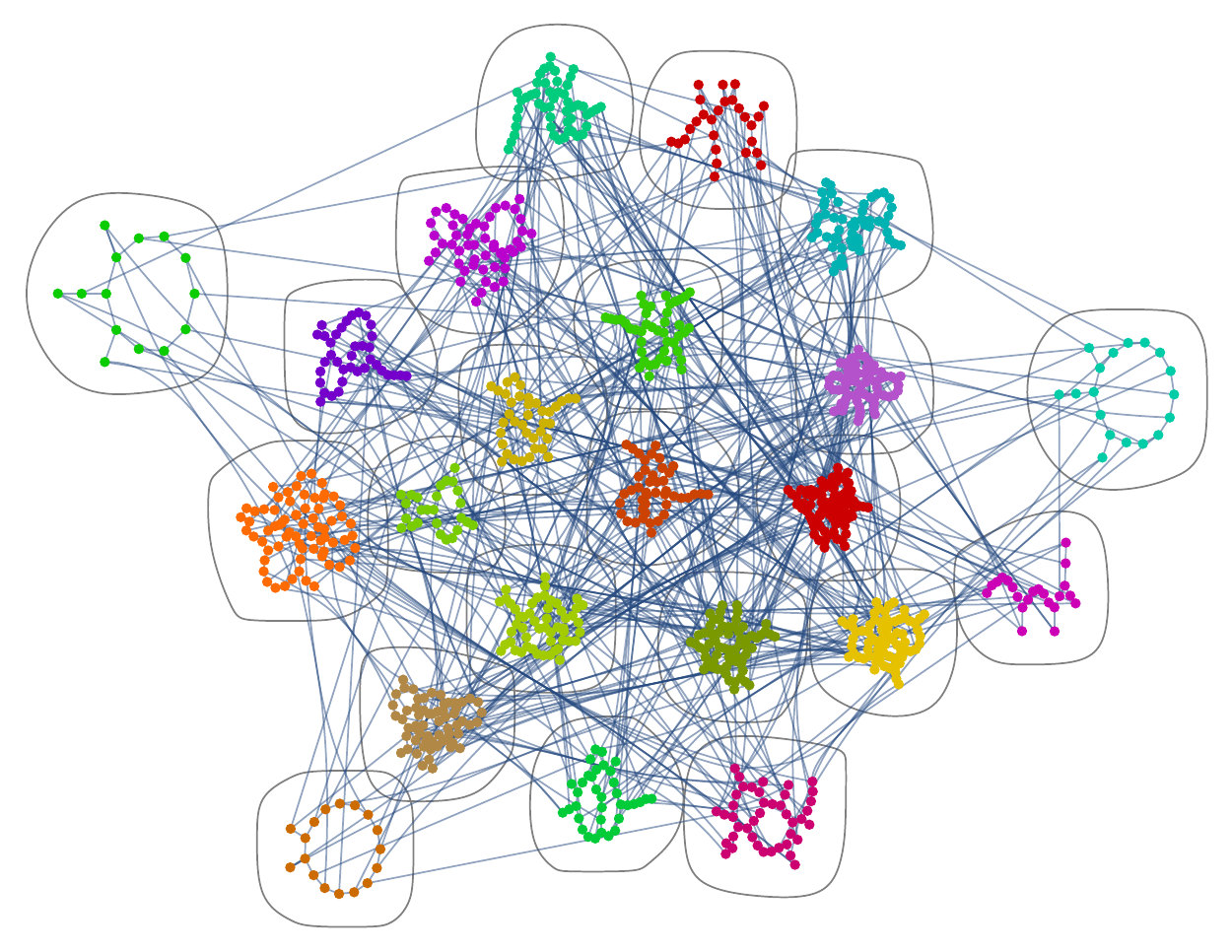}\hfill
\includegraphics*[width=0.5\linewidth]{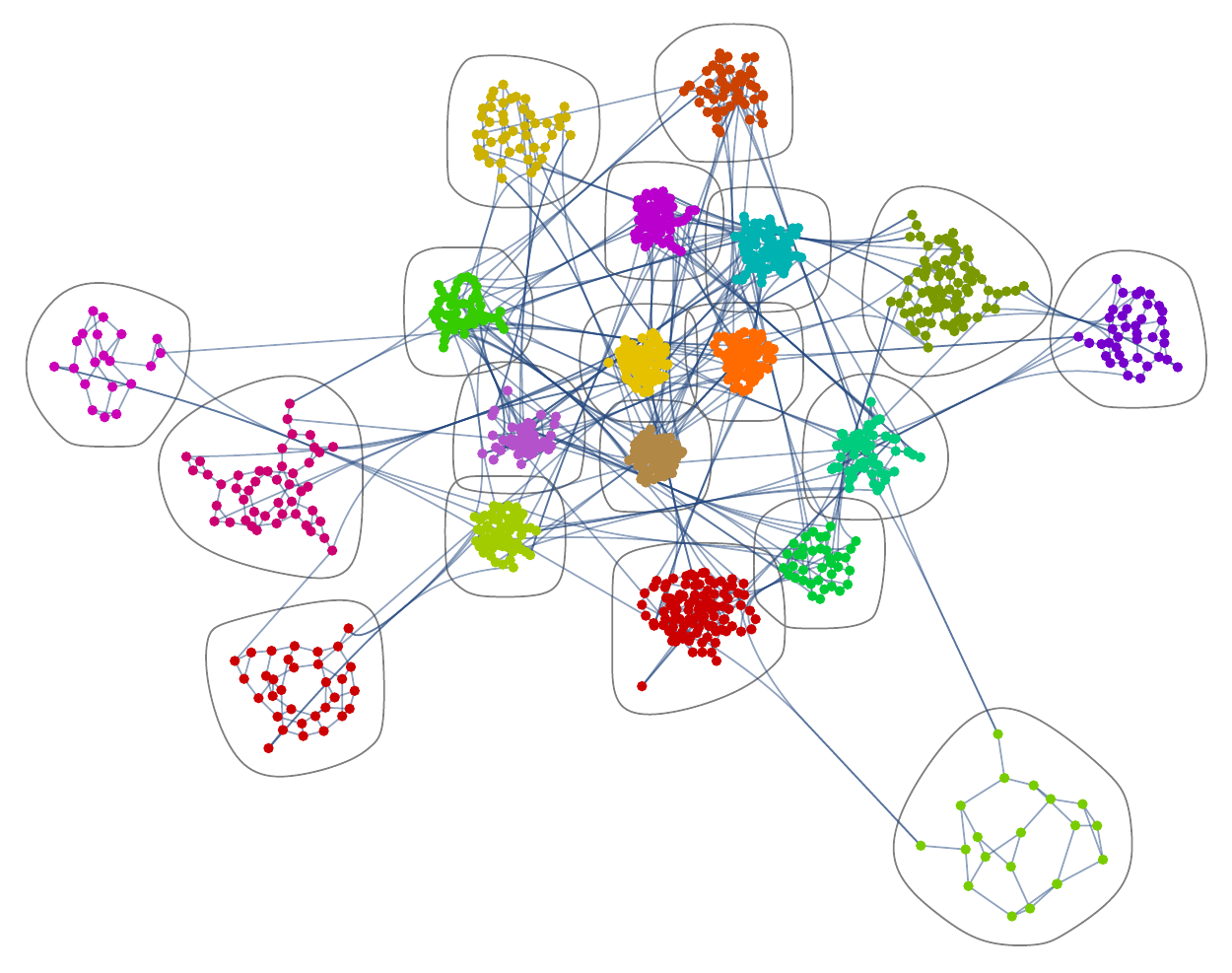}
\caption{\textbf{Community graph plot of a generic RRG of size $N=1000$ (left) and a rewired RRG $\mathcal{R}_{75}$ (right)}.
\label{communautesbis}}
\end{center}
\end{figure}

Below, we first examine the community structure of the Voronoi and mouse networks, and %will 
subsequently discuss how to incorporate these findings into the rewired RRG model. The community structures of the Voronoi and mouse networks are displayed in Fig.~\ref{communautes}. Consistent with a high modularity value, the number of edges within communities is much larger than the number of edges connecting different communities. Moreover, these communities are highly segregated in space, i.e.~with few overlap, as displayed in Fig.~\ref{spheresCommunitiessouris}. In this figure, we use the spatial information about vertices of our 3D graphs to represent each community by a sphere, centered at the barycenter of its vertices and with a radius equal to the standard deviation of vertex positions around that barycenter. Not surprisingly, this suggests that in such systems, communities are a manifestation of the spatial organization of the network, and that vertices in a given spatial region make extensive connections with the neighboring vertices. By contrast, in pure RRGs, the community structure is mainly irrelevant. Communities in RRG graphs are highly connected to each other  (Fig.~\ref{communautesbis}, left) and the average modularity is much lower ($\mu_\textrm{RRG}\sim 0.661$), even if, as said above, the maximization procedure always produces some spurious community structure by construction.  

%One expects that in a biological network embedded in a three-dimensional organ, communities should have a geographical basis to have any meaning.

%We use the optimization method based on modularity to find the communities in the various graphs we study (see Appendix~\ref{appcommunities} for details). The optimization procedure always produces a community structure, even when the substructures have so many links between each other that they are not very relevant to the system.  This relevance can be estimated by a quantity called modularity also produced by the procedure.  

%In Fig.~\ref{communautes} we show how our graphs are organized in communities. The real-life mouse network has visually a clear community structure, as well as the phenomenological Voronoi network. 

%These findings are corroborated in Fig.~\ref{spheresCommunitiessouris}, which shows the spatial distribution of these communities. We relate them with the 3D structure of the graph, assigning to each vertex its spatial postion, which is available for the Voronoi and mouse graphs. As shown in Fig.~\ref{spheresCommunitiessouris}, communities are to a large extend geographical, underlining that in such systems communities are a manifestation of the spatial organization of the graph. One expects that in a biological network embedded in a three-dimensional organ, communities should have a geographical basis to have any meaning.

\begin{figure}[!t]
\begin{center}
\includegraphics*[width=0.99\linewidth]{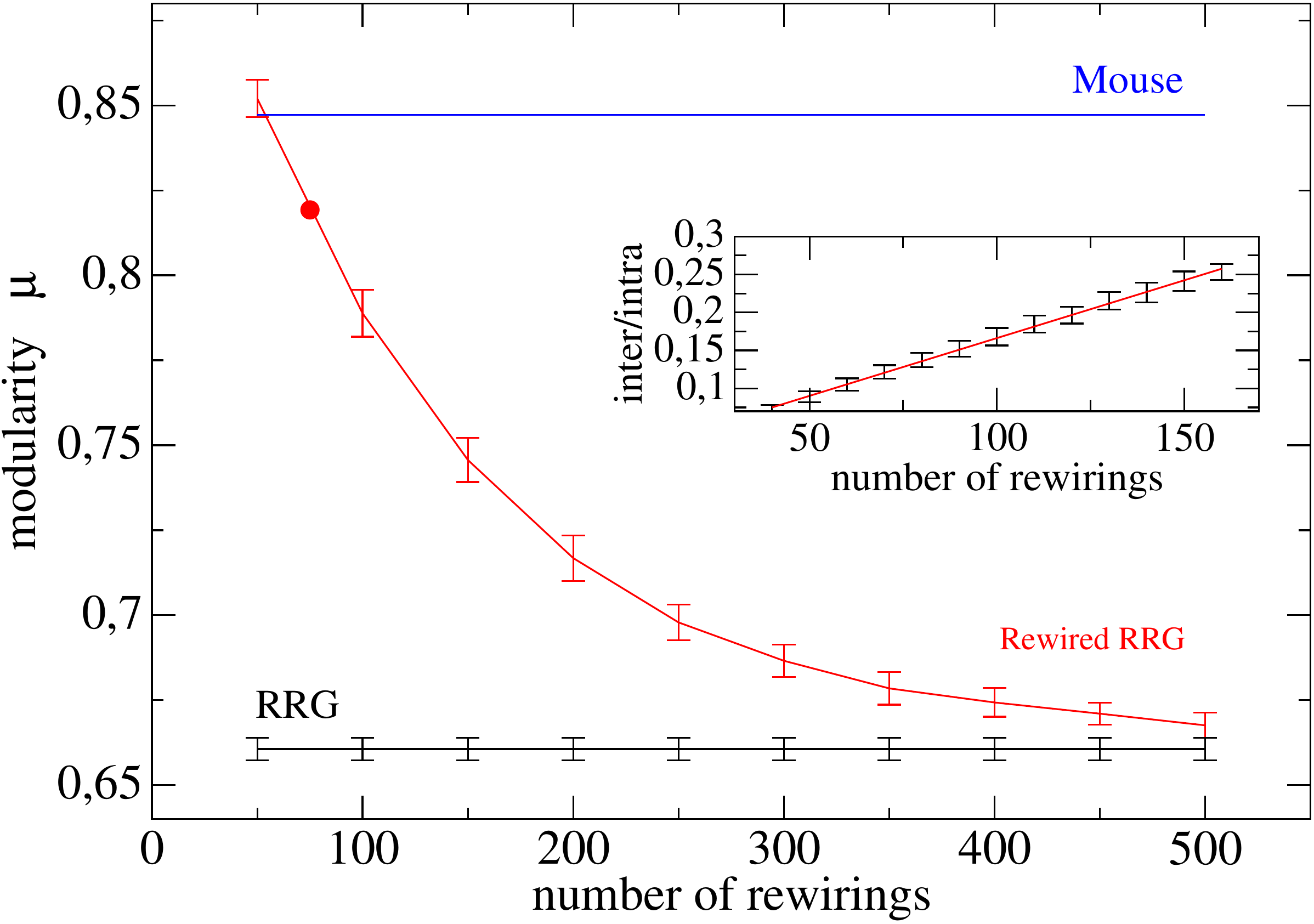}
\caption{\textbf{Graph modularity for rewired RRGs as a function of the number $k$ of rewirings}. Red: rewired RRGs $\mathcal{R}_k$, as defined in Appendix~\ref{construction}, average and standard deviation obtained from 100 realizations \SL{(the red dot indicates 75 rewirings)}; Black: result for generic RRGs of same size ($\mu=0.661$  $\pm$ 0.003 for 100 realizations \SL{of size $N=1146$}); Blue: result for the mouse graph with one single realization.
Voronoi graph $\mathcal{V}$ has modularity $\mu=0.827$, corresponding to $\sim 70$ rewirings.
\textbf{Inset: Mean number of inter- versus intra-community edges in a rewired RRG $\mathcal{R}_k$ as a function of number $k$ of rewirings}. Average and standard deviation from 100 realizations. The red line is the best linear fit ($y=0.0285 + 0.00304 x$). %Voronoi graph $\mathcal{V}$ has 1422 intra links and 182 inter links, thus a ratio  inter/intra=0.128. This corresponds (via the fit) to $74.79\sim 75$ rewirings. \SLNB{reprendre caption (basculer dans le texte les résultats pour le voronoi ?}
\label{meaninterintra}}
\end{center}
\end{figure}

Relevant communities are recovered for the rewired RRGs (see e.g.~Fig.~\ref{communautesbis}, right, where communities associated to rewired RRGs of heterogeneous sizes are displayed). 
From the results presented in the previous Subsection, we expect that increasing the number of rewirings will result in approaching the behavior of a single, larger, RRG. This results in weaker communities, i.e.~a network with lower modularity, as illustrated in Fig.~\ref{meaninterintra}, and \SLNB{an} increased ratio of inter- versus intra-community edges. Below, we seek to estimate the number of rewirings needed to construct heterogeneous rewired RRGs with both community structures (defined by their number and size) and modularity matching those of a given real-life network.

%We now use these tools on our RRG models. As said above, the optimization procedure we use always produces some community structure by construction. However, while communities in a RRG graph (Fig.~\ref{communautesbis} left) are highly linked to each other, and then this community structure is not very relevant, in rewired graphs (Fig.~\ref{communautesbis} right) connections between communities are much weaker, making them more relevant. 

%This can be made more precise by examining the role of rewiring on the ratio between the number of edges that connect vertices within the same community (intralinks) or between two communities (interlinks). This is shown in Fig.~\ref{meaninterintra}. 

For example, in the case of Voronoi $\mathcal{V}$, the network has 1422 intra-community edges and 182 inter-community edges, and thus the ratio between inter and intra edges is $\approx0.128$. Choosing a rewired RRG with elementary RRG sizes given by the communities of $\mathcal{V}$, we find a linear dependence of this ratio with the number of rewirings (inset of Fig.~\ref{meaninterintra}). To match the inter- over intra-edge ratio of $\mathcal{V}$, this yields, via the fit of Fig.~\ref{meaninterintra}, $n_r=74.79\approx 75$ rewirings. 
The so obtained random graphs will be denoted by $\mathcal{R}_{75}$ in the remaining of this paper. Their average modularity (obtained by averaging over 100 realizations) is equal to $\mu_{\mathcal{R}_{75}} \simeq 0.817\pm 0.0065$, which is indeed very close to the modularity of the Voronoi and mouse graphs \footnote{Note that it would also have been possible to choose the number of rewirings that matches the modularity $\mu_\mathcal{V}\sim$0.827 of the Voronoi graph. From the main panel of  Fig.~\ref{meaninterintra}, this would correspond to taking 70 rewirings.}.
%\textcolor{blue}{compléter et comparer à modularité de V. Préciser s'il s'agit de moyennes, y compris dans la phrase suivante}. \textcolor{red}{J'ai choisi de noter $\mathcal{R}_{75}$ tous les graphes possibles avec 75 rewirings car je pense que dans la suite on fait des moyennes. Vous verrez si c'est clair et si ça vous convient. Je n'ai pas encore répercuté cette notation dans la suite du papier. Je le ferai au fil de l'eau en reprenant les sections à partir de IV B}

%\SLNB{Intégrer : Voronoi graph $\mathcal{V}$ has 1422 intra links and 182 inter links, thus a ratio  inter/intra=0.128. This corresponds (via the fit) to $74.79\sim 75$ rewirings.}

Noteworthy, the resulting distribution of loop lengths $P(L)$ for such a number of rewirings matches well the distribution of loop lengths of Voronoi $\mathcal{V}$ (see Fig.~\ref{looplengths}), despite the fact that the 3D structure is entirely absent from the rewired RRG model. This similarity also manifests in the community structure displayed in Figs.~\ref{communautes} and~\ref{communautesbis}.

%\textcolor{red}{Je proppse de supprimer l'ensemble des paragraphes ci-dessous}

%As said before, the degree to which a graph is divided into communities can be further measured by the modularity (see Appendix~\ref{appcommunities} for details). This quantity lies between $-0.5$ and $1.$, and should be positive for the community structure to have any relevance.
%\begin{equation}
%\mu=\sum_\calc\sum_{i,j\in\calc}\left(a_{ij}-\frac{d_id_j}{2m}\right),
%\end{equation}
%where $a_{ij}$ is equal to 1 if there is an edge connecting vertices $i$ and $j$, and to 0 otherwise, $d_i$ is the number of outgoing links of $i$, $m$ is the total number of edges, and the sum runs over all communities $\calc$ detected by the algorithm. 

%For a pure RRG, we obtain an average modularity $\mu_\textrm{RRG}\sim 0.677$, while for the mouse graph $\mu_\textrm{mouse}\sim 0.8464$. 

%Several realizations of the Voronoi graph show a modularity between $0.82$ and $0.83$. For a RRG rewired graph, the modularily decreases with $n_r$ from $\sim 0.9$ to the generic RRG value which is reached asymptotically as the number of rewirings goes to infinity. \textcolor{red}{homogeneous sub RRGs ? Maybe move earlier (around line 256) ?}

%\textcolor{red}{Suite à garder}

The results of this Section show that topological features such as the community structure are important properties of intracortical microvascular networks. These properties can be implemented in the RRG model by adding a community structure and a certain amount of rewiring. Such models can describe correctly the loop distribution and the modularity of the real networks, while having no other features left in the model. 
They \SL{thus} enable \SLNB{averages over all realizations of these networks, which all have the same structure, to be performed,} a useful tool to get statistically significant quantities and assess the variability of the results. 
\SL{In fact, the rewired RRG ensemble corresponding to a given number of rewirings $\mathcal{R}_k$ is the subset of the RRG ensemble which describes the universal behavior of regular graphs having a given heterogeneous community size distribution and a given community strength. The Voronoi and mouse networks are specific instances of this ensemble and turn out to display topological quantities which are in line with the mean result of the equivalent rewired RRGs.}
%  \0G{The rewired RRG ensembles indeed describe the universal behavior of regular graphs having a given community strength.  The Voronoi and mouse networks, which are specific instances of this ensemble, turnout to display topological quantities which are in line with the mean result of the equivalent rewired RRG.}
 We will use these models in the next Section to study the resilience of the above networks to single-edge occlusion and assess its dependence on modularity.

%\SLNB{resilience ? vulnerability}
%\SLNB{response}

%%%%%%%%%%%%%%%%%%%%%%%%%%%%%%%%%%%%%%%%%%%%%%%%%%%%%%%%%%%%%%%%%%%%%%%%%
%%%%%%%%%%%%%%%%%%%%%%%%%%%%%%%%%%%%%%%%%%%%%%%%%%%%%%%%%%%%%%%%%%%%%%%%%
\section{Blood flow through the network}\label{secflux}
%%%%%%%%%%%%%%%%%%%%%%%%%%%%%%%%%%%%%%%%%%%%%%%%%%%%%%%%%%%%%%%%%%%%%%%%%
%%%%%%%%%%%%%%%%%%%%%%%%%%%%%%%%%%%%%%%%%%%%%%%%%%%%%%%%%%%%%%%%%%%%%%%%%
We now connect the previous graphs with the outer world by adding a single inlet edge and a single outlet edge (see Fig.~\ref{fleches}). 
Following~\cite{Hudetz1993,Pozrikidis2012,gavrilchenko_resilience_2019}, we also ignore the %Because we ignore the 
complex rheology of blood, so that imposing a constant pressure (potential) difference between the tips of these edges results in establishing a stationary flow field throughout the network, that can be obtained by inverting a linear system of equations~\cite{Pozrikidis2012,Lorthois_Book_2019}. In the present Section, we shall examine how occluding a single vessel affects the total flow rate transported through the network.%, \OB{focusing on the simplest case where there are only one injection point and one outlet point in the network (see Fig.~\ref{fleches}).}

%%%%%%%%%%%%%%%%%%%%%%%%%%%%%%%%%%%%%%%%%%%%%%%%%%%%%%%%%%%%%%%%%%%%%%%%%
\subsection{Definitions}

%%%%%%%%%%%%%%%%%%%%%%%%%%%%%%%%%%%%%%%%%%%%%%%%%%%%%%%%%%%%%%%%%%%%%%%%%

%For a graph $\cg$ with $M$ vertices \SL{isn't it $N$ here ?} let $p_i$ be  the  potential (pressure) at vertex i for $1\leq i\leq M$. The local flux (flow rate) from $i$ to $j$ is defined as
%\begin{equation}
%\label{qij}
%q_{ij}=\gamma_{ij}(p_i-p_j),
%\end{equation}
%where $\gamma$ is the matrix of conductivities. In the present paper, we restrict ourselves for simplicity to the case where a single inlet edge and a single outlet edge only are added to the graph. In this case, by mass conservation, the total flow rate transported through the network is equal to the inlet and outlet fluxes.

Starting from a graph $\cg$ with $M=N-2$ vertices, we pick up two edges at random, and add two vertices $I$ and $O$ in the middle of these edges. To these vertices, which now are the $N-1$$^{th}$ and $N$$^{th}$ vertices of the graph, we respectively connect an inlet vertex $I'$ and an outlet vertex $O'$. %, where we impose given potentials $p_{I'}$ and $p_{0'}$.
Note that, in that way, a graph with constant connectivity $z=3$ keeps that property.

%For a graph $\cg$ with $M$ vertices \SL{isn't it $N$ here ?} 
Let $p_i$ denote the  potential (pressure) at vertex i for $1\leq i\leq N$. The local flux (flow rate) from $i$ to $j$ is defined as
\begin{equation}
\label{qij}
q_{ij}=\gamma_{ij}(p_i-p_j),
\end{equation}
where $\gamma$ is the matrix of conductivities. %In the present paper, we restrict ourselves for simplicity to the case where a single inlet edge and a single outlet edge only are added to the graph. %In this case, by mass conservation, the total flow rate transported through the network is equal to the inlet and outlet fluxes.

For $i=1$ to $N-2$, Kirchhoff's current law yields
\begin{equation}
\label{kirchhoff}
\sum_{j}\gamma_{ij}p_j-\left(\sum_{ j}\gamma_{ij}\right)p_i=0,\qquad 1\leq i\leq N-2,
\end{equation}
where $\gamma_{ij}$ is the conductivity of edge $ij$.
For $i=N-1$ or $N$, there is an additional flux $\gamma_{i i'}(p_{i}-p_{i'})$, where $p_{i'}$ is the  corresponding imposed potential ($p_{I'}$ or $p_{0'}$), yielding
\begin{equation}
\label{defb}
\sum_{j\neq i'}\gamma_{ij}p_j-\left(\sum_{j}\gamma_{ij}\right)p_i=-\gamma_{ii'}p_{i'},\qquad i=N-1,N
\end{equation}
where $\gamma_{ii'}$ is the conductivity of the newly added edges, which will be denoted by  $\gamma_{I'}$ for the inlet and $\gamma_{O'}$ for the outlet in the following. 
Let $A$ be the $N\times N$ matrix defined by
\begin{equation}
\label{mij}
A_{ij}=\left\{
\begin{array}{ll}\gamma_{ij}&\qquad \textrm{if } i\neq j,\\ 
-\sum_k \gamma_{ik}&\qquad \textrm{if } i=j,
\end{array} 
\right.
\end{equation}
for $1\leq i,j\leq N$, and the sum over $k$ runs over all neighbours of $i$, including $i'$. \SL{Matrix $-A$ is known as the (weighted) graph Laplacian matrix, and its spectral properties have been extensively studied \cite{harary66, fiedler73}}. We define vector $b$ as the vector with entries $-\gamma_{i i'}p_{i'}$ at position $i=N-1$ and $N$ and 0 elsewhere, that is, $b=(0,\ldots,0,-\gamma_{I'}p_{I'},-\gamma_{O'} p_{O'})^T$. Then, from Eqs.~\eqref{kirchhoff} and \eqref{defb}, the vector $p$ is the unique solution of $Ap=b$.
Generically, $A$ is invertible \SL{(note that the sum over $k$ in \eqref{mij} includes $i'$)}, so that $p$ is explicitly given by $p=A^{-1}b$. 

\begin{figure}[!t]
\begin{center}
\includegraphics*[width=0.92\linewidth]{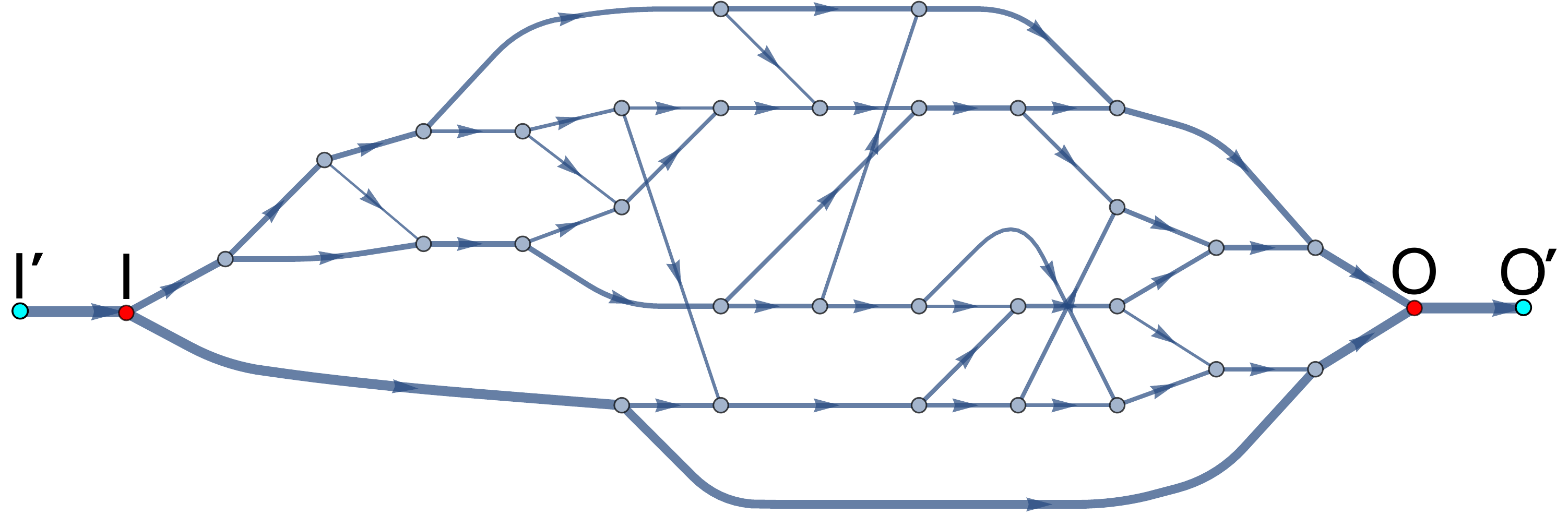}
\caption{\textbf{Flow field in a single RRG \SL{with $N=32$ vertices}}. The two boundary vertices ($I'$ and $O'$) are highlighted in cyan. The two vertices added to connect these boundaries to the graph ($I$ and $O$), are highlighted in red. Arrows show the direction of flow $q$ while edge thickness is proportional to $1/\log |q|$.}
\label{fleches}
\end{center}
\end{figure}

In Fig.~\ref{fleches}, we illustrate the fluxes $q_{ij}$ for one inlet point $I'$ on the left and one outlet point $O'$ on the right of a single RRG. By mass conservation, the total flow rate $Q$ transported through the network is equal to the inlet and outlet fluxes. Thus, with the above notations,  $Q=\gamma_{I'}(p_{I'}-p_I)=\gamma_{O'}(p_{O}-p_{O'})=\Gamma(p_{I'}-p_{O'})$, where $\Gamma$ denotes the overall network conductance.
We also denote by $\delta P=p_{I'}-p_{O'}$ the fixed value of the potential difference between the inlet and outlet. Then $Q=\Gamma\delta P$. We will now examine how $Q$, or equivalently $\Gamma$, is affected by the removal of one edge.

%%%%%%%%%%%%%%%%%%%%%%%%%%%%%%%%%%%%%%%%%%%%%%%%%%%%%%%%%%%%%%%%%%%%%%%%%
\subsection{Flow reduction induced by removing one edge}\label{remove}
%%%%%%%%%%%%%%%%%%%%%%%%%%%%%%%%%%%%%%%%%%%%%%%%%%%%%%%%%%%%%%%%%%%%%%%%%

%As written above, by mass conservation, the total flow rate $Q$ transported through the network is equal to the inlet and outlet fluxes. Thus, with the above notations,  $Q=p_{I'}-p_I=p_{O}-p_{O'}$.

We now denote by $Q_{ij}'$ the total flow rate transported through the network when edge $i-j$ is removed, and by $\delta Q_{ij}=Q-Q_{ij}'$ the corresponding flow reduction. This flow reduction can be expressed as follows (see Appendix~\ref{removeone})

\begin{equation}
\label{qqpgeneral_dim}
%\delta Q_{ij}=\frac{q_{ij}^2}{t_{ij}},
%\end{equation}
%where $q_{ij}$ is the initial flux in the removed edge and $t_{ij}=1+(A^{-1})_{ii}+(A^{-1})_{jj}-2(A^{-1})_{ij}$ only depends on the topology of the graph. 
\frac{\delta Q_{ij}}{Q} \equiv\frac{(Q-Q'_{ij})}{Q}=\frac{\Gamma}{\gamma_{ij}}\frac{1}{t_{ij}}\left(\frac{q_{ij}}{Q}\right)^2\,,
\end{equation}
where $q_{ij}$ is the initial flux in the removed edge %, $\Gamma$ is the overall network conductance 
and $t_{ij}=1+\gamma_{ij}[(A^{-1})_{ii}+(A^{-1})_{jj}-2(A^{-1})_{ij}]$ 
%$t_{ij}=1+(A^{-1})_{ii}+(A^{-1})_{jj}-2(A^{-1})_{ij}$ 
is non-dimensional. 

In what follows, for simplicity, we will only consider the case where the pressure reduction $\delta P=p_{I'}-p_{O'}=1$ is fixed, 
so that $Q=\Gamma$, and where all conductivities $\gamma_{ij}$ are taken equal to 1.
 %When removing edge $i-j$ with flux $q_{ij}$, the global flux becomes $Q'$. One can calculate the corresponding drop in global flux $Q-Q'=\delta Q_{ij}$ (see Appendix~\ref{removeone}). It is given by
Equation \eqref{qqpgeneral_dim} reduces to
\begin{equation}
\label{qqpgeneral}
\delta Q_{ij}=\frac{q_{ij}^2}{t_{ij}}\,.
\end{equation}
Equation \eqref{qqpgeneral} states that the total flow reduction induced by removing one edge $i-j$ is quadratic in the initial flux $q_{ij}$ through this edge, and inversely proportional to $t_{ij}$. 

\SL{Noteworthy, the term $t_{ij}$ in Eq.~\eqref{qqpgeneral} is analogous to expressions known as Line Outage Distribution Factors (LODF) describing the response to outage of an edge in an electric power grid~\cite{pgoc84}. Similar expressions have also been derived in the broader context of network theory~\cite{Man17}, including for the study of single edge removal in space-filling model networks of biological vasculature~\cite{gavrilchenko_resilience_2019}.}

%\SL{Equation \eqref{qqpgeneral} is analogous to expressions known as Line Outage Distribution Factors (LODF) describing the response to outage of an edge in an electric power grid, see Eq.~11A.13 of \cite{pgoc84}.
%Similar formulas were obtained in the broader context of network theory \cite{Man17}.}

Below, we further examine the properties of these two quantities, \SL{$t_{ij}$ and $q_{ij}$}, governing the flow reduction.

%the reduction of the total flux $Q$ is quadratic in the flux $q$ removed. The prefactor by which $q^2$ has to be multiplied only depends on the structure of the graph.

%%%%%%%%%%%%%%%%%%%%%%%%%%%%%%%%%%%%%%%%%%%%%%%%%%%%%%%%%%%%%%%%%%%%%%%%%
\subsection{Cauchy laws for the quantities governing the flow reduction}
%%%%%%%%%%%%%%%%%%%%%%%%%%%%%%%%%%%%%%%%%%%%%%%%%%%%%%%%%%%%%%%%%%%%%%%%%

\begin{figure}[!t]
\begin{center}
\includegraphics*[width=.99\linewidth]{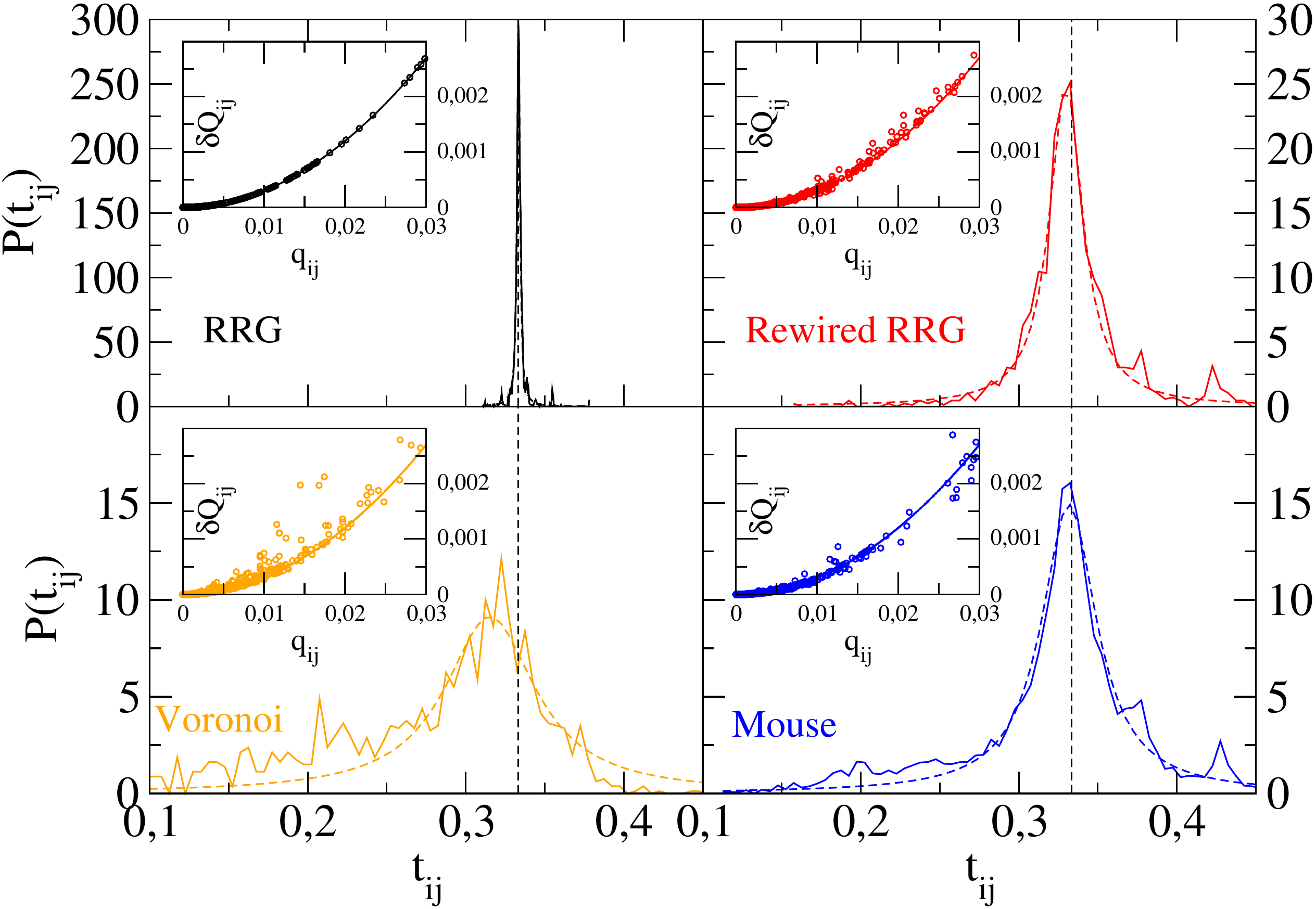}\caption{\textbf{Distribution \SL{of $t_{ij}$} for $(i,j)$ running over all edges of a single realization of a graph}. Solid lines display numerical results, dashed lines Cauchy fits (Eq.~\eqref{cauchylaw}). Top: RRG \SL{with $N=1000$} (left), rewired RRG $\mathcal{R}_{75}$ (right); Bottom: Voronoi $\mathcal{V}$ (left), Mouse (right). \textbf{Insets: flow reduction $\delta Q_{ij}$ as a function of $q_{ij}$, illustrating the quadratic dependence given by Eq.~\eqref{qqpgeneral}}.}
\label{figcauchy}
\end{center}
\end{figure}

It is first possible to get an insight into the denominator $t_{ij}$ in Eq.~\eqref{qqpgeneral} by considering the asymptotic case of very large regular trees (see e.g.~\cite{Abou73,BogGir13}). %In this case, $A=\mathcal{A}-z\mathbb{1}$, where $\mathbb{1}$ is the identity matrix. 
For that purpose, let us denote by $G(x)=(A-x\mathbb{1})^{-1}$ the Green function of matrix $A$, so that $A^{-1}=G(0)$. In the case where the off-diagonal elements of $A$ are those of the adjacency matrix of an infinite regular tree, one can find a recursive relation between the diagonal elements of $G(x)$. The latter expresses $G_{ii}(x)$ as a function of the $G_{jj}(x)$, with $j$ the children of $i$, and of the diagonal elements of $A$ (see e.g.~\cite{Abou73,BogGir13}). For infinite trees, all vertices $i$ are at the root of statistically identical trees. Therefore $G_{ii}(x)$ behaves as a random variable $G$ and all $G_{jj}(x)$ behave as independent random variables $G_p$ with the same distribution as $G$. These are solutions of the recursive equation
\begin{equation}
\label{treeeq}
G=\frac{1}{c-x-\sum_{p=1}^{z-1}G_p}\,,
\end{equation}
where $z$ is the graph connectivity and $c$ is a random variable distributed in the same way as the diagonal elements of $A$. When the graph is no longer an infinite tree but a finite random regular graph, Eq.~\eqref{treeeq} is only approximate.  

It is known in statistical physics\SL{~\cite{BogGir13,kraAnnPhys2018}} that the mean-field solution of this recursive equation amounts to approximating the exact solution (in the form of a probability distribution of $G_{ii}(x)$ as a random variable)  by symmetric Cauchy distributions. As a result, $(A^{-1})_{ii}$ is expected to follow a Cauchy distribution, so that, by stability of Cauchy distributions, the quantity $t_{ij}=1+(A^{-1})_{ii}+(A^{-1})_{jj}-2(A^{-1})_{ij}$, which corresponds to the denominator in \eqref{qqpgeneral}, is also \SL{expected} to follow a Cauchy distribution.

In Fig.~\ref{figcauchy}, we plot the distributions of $t_{ij}$ for all our networks models, from RRGs to the mouse network. Whatever the considered graph, these distributions are indeed well-fitted by  Cauchy distributions
\begin{equation}
\label{cauchylaw}
P(t)=\frac{1}{\pi}\frac{a}{a^2+(t-t_0)^2},
\end{equation}
where $t_0$ is the median value and $a$ is the half width at half maximum (HWHM)~\cite{encofmath}. Besides, in all cases, 
we numerically find that the median value $t_0$ is close to $1/3$, and thus the prefactor of $q_{ij}^2$ in \eqref{qqpgeneral} is close to 3.
This result can be recovered by generalizing the reasoning of~\cite{Kirk73,Pozrikidis2012}, initially introduced to study the dilute regime of bond percolation. For regular networks of connectivity $z$ with a single inlet and outlet, we show in Appendix~\ref{meantij} that the flow reduction can be approximated by 
%is consistent with predictions obtained from the dilute regime of bond percolation. Generalizing the reasoning of~\cite{Kirk73,Pozrikidis2012} to regular networks of connectivity $z$ with a single inlet and outlet, we show in Appendix~\ref{meantij} that the flow reduction can be approximated by
\begin{equation}
\label{qqpgeneralz}
\delta Q_{ij}=\frac{z}{z-2}(p_i-p_j)^2
\end{equation}
 in the limit of infinite size and under the assumption that the graph is isotropic at the inlet and outlet.
%It is possible to understand why the median value of the Cauchy law concentrates around $1/3$ using simple physical interpretation for the quantity $t_{ij}$. As we show in Appendix~\ref{meantij}, following the methods in~\cite{Kirk73,Pozrikidis2012},  for a graph with constant connectivity $z$, the flow reduction can be approximated by
%\begin{equation}
%\label{qqpgeneralz}
%\delta Q_{ij}=\frac{z}{z-2}(p_i-p_j)^2
%\end{equation}
 %under the assumption that the graph is isotropic at the injection point and in the limit where its size goes to infinity. 
  The inset of Fig.~\ref{figcauchy} shows that this quadratic law is well verified whatever the considered graph. In our case, $z=3$ and thus the prefactor of $(p_i-p_j)^2$ in \eqref{qqpgeneralz} is  $\frac{z}{z-2}=3$, recovering $1/3$ as the median value $t_0$ of the Cauchy distribution for $t_{ij}$. 
 
In addition, Cauchy laws have fat tails. As a result, the occurrence probability of small and large values of quantities following Cauchy distributions is much larger than for normal distributions. For instance, the probability of values smaller than $t_0-3a$, where $t_0$ denotes the median and $a$ is the HWHM, is $\approx 0.0002$ for a Gaussian distribution whereas it is $\approx 0.1 $ for a Cauchy distribution, i.e.~five hundred times higher. Thus, there is a number of values of the $t_{ij}$ which are very small. When such an edge $i-j$ is removed, this translates into a prefactor of the flow reduction $q_{ij}$ which is particularly high, inducing a significant drop in the total flow $Q$. 
As a result, the larger the Cauchy distribution in Fig.~\ref{figcauchy}, the larger the fluctuations observed around Eq.~\eqref{qqpgeneralz} in the corresponding inset.

%Attention : SL : gamma n'est pas la standard deviation de la cauchy car il n'y en a pas : diverge. 

%\SLNB{Dominé par les événements rares : est-ce qu'on a une borne supérieure pour les tij ? Matrice d'entiers ? Support borné ?} 
\begin{figure}[!t]
\begin{center}
\includegraphics*[width=.99\linewidth]{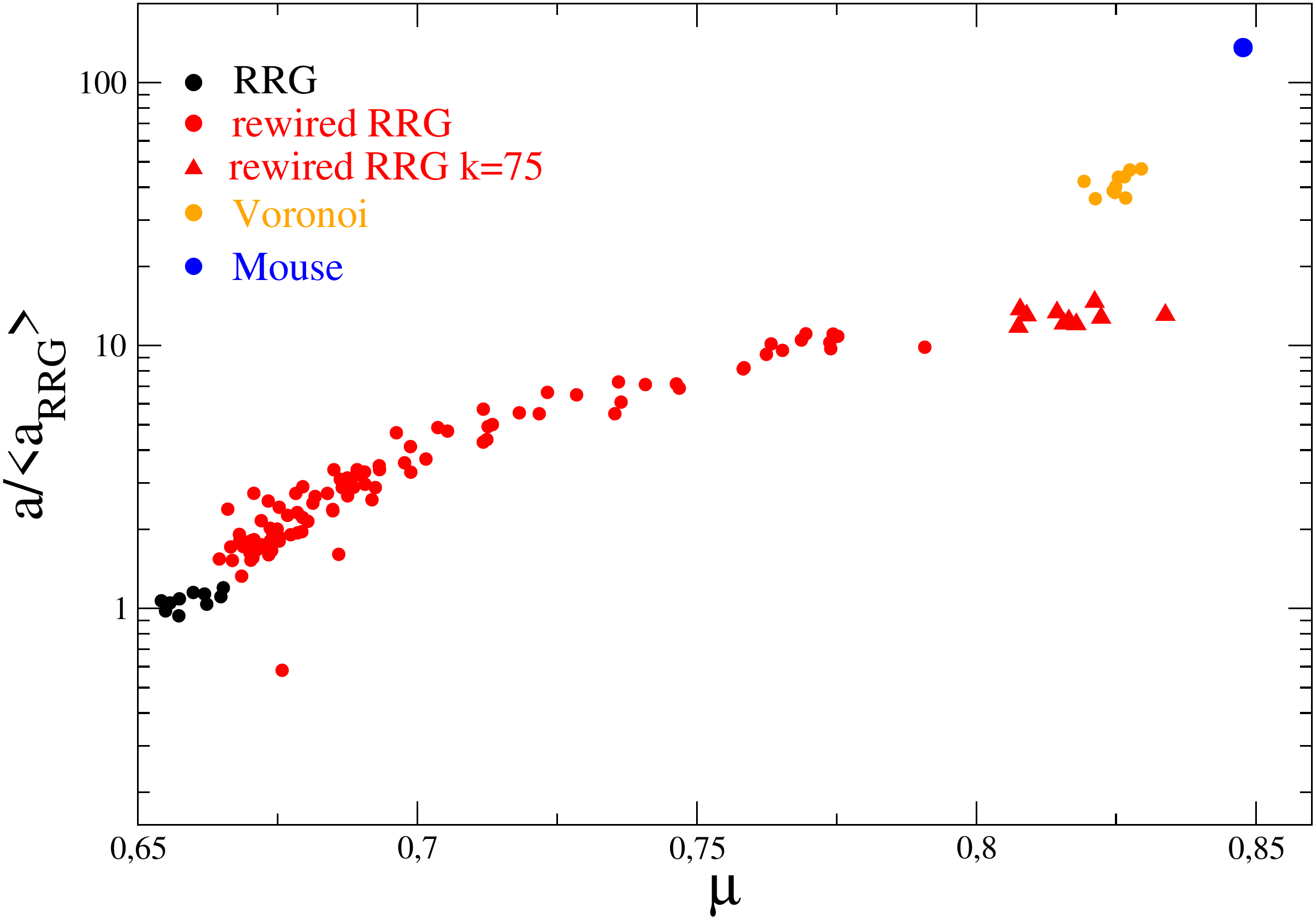}
\caption{\textbf{Width $a$ of the Cauchy distribution \SL{for $t_{ij}$, as defined in Eq.~}\eqref{cauchylaw} (normalized by the average Cauchy width of a RRG graph of same size) as a function of graph modularity}. Each point is a single graph realization. RRG (black), rewired RRG (red) for $\mathcal{R}_{75}$ (triangles) and $\mathcal{R}_k$ with $k=100$ to 200 (circles) rewirings, Voronoi (orange), and Mouse (blue).\label{figmodcau}}
\end{center}
\end{figure}

%It is possible to understand why the median value of the Cauchy law concentrates around $1/3$ using simple physical interpretation for the quantity $t_{ij}$. As we show in Appendix~\ref{meantij}, following the methods in~\cite{Kirk73},  for a graph with constant connectivity $z$, the flow reduction can be approximated by
%\begin{equation}
%\label{qqpgeneralz}
%\delta Q=\frac{z}{z-2}(p_1-p_2)^2
%\end{equation}
 %under the assumption that the graph is isotropic at the injection point and in the limit where its size goes to infinity. The inset of Fig.~\ref{figcauchy} show that this quadratic law is well verified. In our case $z=3$ and thus the prefactor of $(p_1-p_2)^2$ in \eqref{qqpgeneralz} is indeed $\frac{z-2}{z}=\frac{1}{3}$, implying that the median is $\langle t_{ij} \rangle =1/3$.

Importantly, the half width $a$ of the Cauchy distribution strongly depends on the community structure of the underlying graph. In Fig.~\ref{figmodcau}, we show that for graphs of comparable size, the width increases with the graph modularity, significantly increasing the probability of large flow reductions resulting from the removal of a single edge.
\SL{Interestingly, while the Voronoi and mouse networks follow the same trend as the rewired RRGs $\mathcal{R}_k$, the width  of their Cauchy distribution is larger than that obtained for rewired RRGs, up to an order of magnitude for the mouse network. It is unlikely that this increase can be due to the random dispersion of results in the rewired RRG ensemble. Rather, in the same way that the $\mathcal{R}_k$s, despite being specific realizations of graphs which belong to the RRG ensemble, display on average higher modularities than more probable realizations directly drawn from the RRG ensemble (compare modularity of red versus black symbols in Fig.~\ref{figmodcau}), we may expect that other topological peculiarities of the community structure in the Voronoi and mouse networks (e.g. heterogeneous distribution of community size, increased probability of inter-links connecting neighbor communities in the 3D space) might also systematically bias the width of the Cauchy distribution. Additional biological data are probably needed to explore this point.}%\OG{We note that the Voronoi and mouse networks, although they follow the same trend, are above the average value of rewired RRGs. As these networks are just specific instances of the general ensembles of rewired RRGs, this increase can be due to the random dispersion of results in the rewired RRGs, or to a systematic effect due to the peculiarities of the community structure in vascular networks. Additional biological data are probably needed to settle this point.}

%Rev 1 : However, it seems that modularity alone does not fully capture the differences in their quantities, and especially the very large jumps observed between the different network types. Can the authors comment on the reasons these jumps are observed?}

%Rev 2 : These deviations are quite large (about an order of magnitude between red and blue for nearly the same modularity), and hint at other fundamental topological contributors. Is there any approximation that can be made by some expansion of tij to more explicitly get at the main contribution of A to tij? If not, I would like to see some analysis and discussion explaining some of this discrepancy. One example of a potential analysis would be to find some other networkmeasure (number of nodes, higher-order topology, etc.) that, when used as a regressor in relating a to $\mu$, improves the fit of all of the points of Figure 9 using one curve.

% As increasing the width of the distribution of $t_{ij}$ makes more probable very small values, this correspondingly, through \eqref{qqpgeneral}, increases the probability of large flux variations. 

\begin{figure}[!t]
\begin{center}
\includegraphics*[width=.99\linewidth]{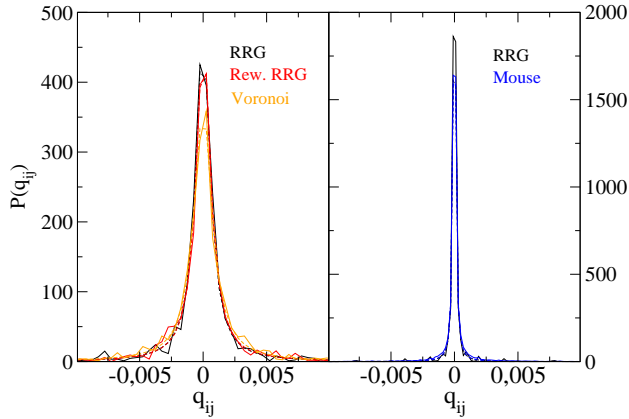}\caption{\SL{\textbf{Distribution of $q_{ij}$ for $(i,j)$ running over all edges of a single realization of a graph}. Same color code as in Fig.~\ref{figcauchy}.  Left~($N$=1176): $a=6.9\times10^{-4}$ for RRG; $a=7.0\times 10^{-4}$ for rewired RRG; $a=8.8\times10^{-4}$ for Voronoi. Right~($N$=8720): $a=9.9\times10^{-5}$ for RRG; $a=1.05\times10^{-4}$ for mouse.}}
\label{figcauchy_qij}
\end{center}
\end{figure}

We now turn to the distribution of $q_{ij}$, the numerator in \eqref{qqpgeneral}. This distribution can be understood based on similar arguments as above. From \eqref{qij}, local fluxes are indeed obtained by linear combinations of potentials $p=A^{-1}b$, leading to a Cauchy distribution in the case of infinite trees. Besides, because of  flow reversibility, its median value is expected to be zero. This is consistent with the blood flow distributions computed in realistic intracortical networks from mice~\cite{goirand}. These distributions have fat algebraic tails (power laws with exponent -2) characteristic of Cauchy distributions, which can alternately be interpreted, based on hydrodynamic arguments, as emerging from dipole-driven flows on random networks. Their properties have been investigated in~\cite{goirand}, but the impact of the community structure has not been considered. Following the same approach as above, we find however that\SL{, for graphs of comparable size,} modularity affects significantly less the width of the distribution of $q_{ij}$ compared to the one of $t_{ij}$ \SL{(see Fig.~\ref{figcauchy_qij})}.

%that in the case of the  $q_{ij}$, the effect of the modularity on the width of the distribution is much less pronounced than for the  $t_{ij}$}. 

%As concerns the distribution of $q_{ij}$, the numerator in \eqref{qqpgeneral}, the same argument from statistical physics as for the $t_{ij}$ (see above) also leads to a Cauchy law; this could be also obtained through hydrodynamical arguments~\cite{goirand}, and the numerical distributions~\cite{goirand} indeed have the algebraic tails characteristic of Cauchy distributions.

%\begin{eqnarray}
%p(q) \sim \frac{1}{1+\left(\frac{q}{q_c}\right)^2}
%\label{FlowDistrib}
%\end{eqnarray}

%%%%%%%%%%%%%%%%%%%%%%%%%%%%%%%%%%%%%%%%%%%%%%%%%%%%%%%%%%%%%%%%%%%%%%%%%
\subsection{Distribution of flow reductions}
%%%%%%%%%%%%%%%%%%%%%%%%%%%%%%%%%%%%%%%%%%%%%%%%%%%%%%%%%%%%%%%%%%%%%%%%%

After having determined the distribution of the denominators and numerators in Eq.~\eqref{qqpgeneral}, we now turn to the distribution of flow reductions $\delta Q_{ij}$, which controls the variability of the global network response to the removal of a single edge. There is no general result which describes this distribution for both $q_{ij}$ and $t_{ij}$ following Cauchy distributions. In Appendix~\ref{removal}, however, assuming that the  distribution of flow reductions is mainly controlled by the distribution of flow rates within the  network, we show that for a regular graph and in the limit of large sizes, the left tail of the distribution of $\ln \delta Q_{ij}$ is expected to follow $P(\ln \delta Q)\sim 1/\delta Q^{-1/2}$ and the right tail is expected to follow $P(\ln \delta Q)\sim 1/\delta Q^{1/2}$.

\begin{figure}[!t]
\begin{center}
\includegraphics*[width=.89\linewidth]{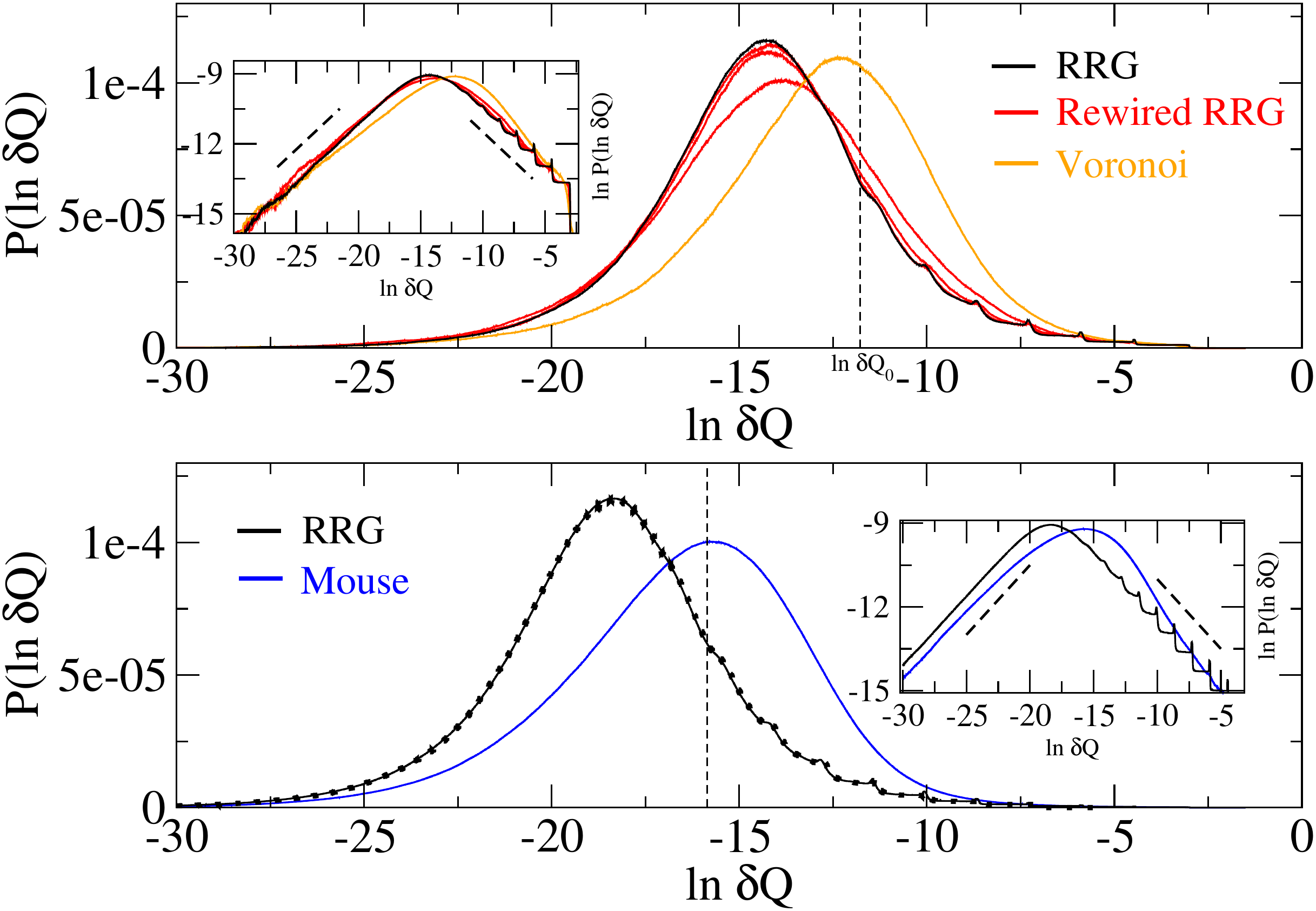}\\
\caption{\textbf{Top panel:
Distribution of $\ln\delta Q$ for $(i,j)$ over all edges.} Black : 100 realizations of generic RRGs with $N_1=1176$; Orange: Voronoi (100 realizations); Red: 1000 realizations of rewired 
RRGs $\mathcal{R}_k$ with random $k$, $100\leq k\leq 600$, divided into 20 sets of 50 realizations, where in each set graphs have approximately the same modularity; histograms for sets 10, 15 and 20 are shown (three red curves with modularity increasing from left to right). All histograms are averaged over windows of size $\sim 1$ in $\ln\delta Q$. The dashed vertical line indicates position $\ln\delta Q_0=-11.773$ where $25\%$ of the probability lies on the right for the generic RRG of size $N_1$. 
\textbf{Bottom panel: same for mouse (blue) and RRGs with $N_2=8720$ (black, 100 realizations)} ; dotted black corresponds to the RRG of size $N_1$ translated to the left by $\ln \frac{N_2^2}{N_1^2}\simeq 4$ units in logarithmic scale. With this translation, both RRG distributions coincide, showing that the $1/N^2$ scaling obtained in Appendix E corrects perfectly for the size effect in case of RRGs. The dashed vertical line corresponds to the vertical line in the top panel translated in the same way.  \textbf{Insets: same in log scale}. Dashed lines indicate slopes $\pm\frac12$. \label{figdeltaq}}
\end{center}
\end{figure}

\begin{figure}[!t]
\begin{center}
\includegraphics*[width=.89\linewidth]{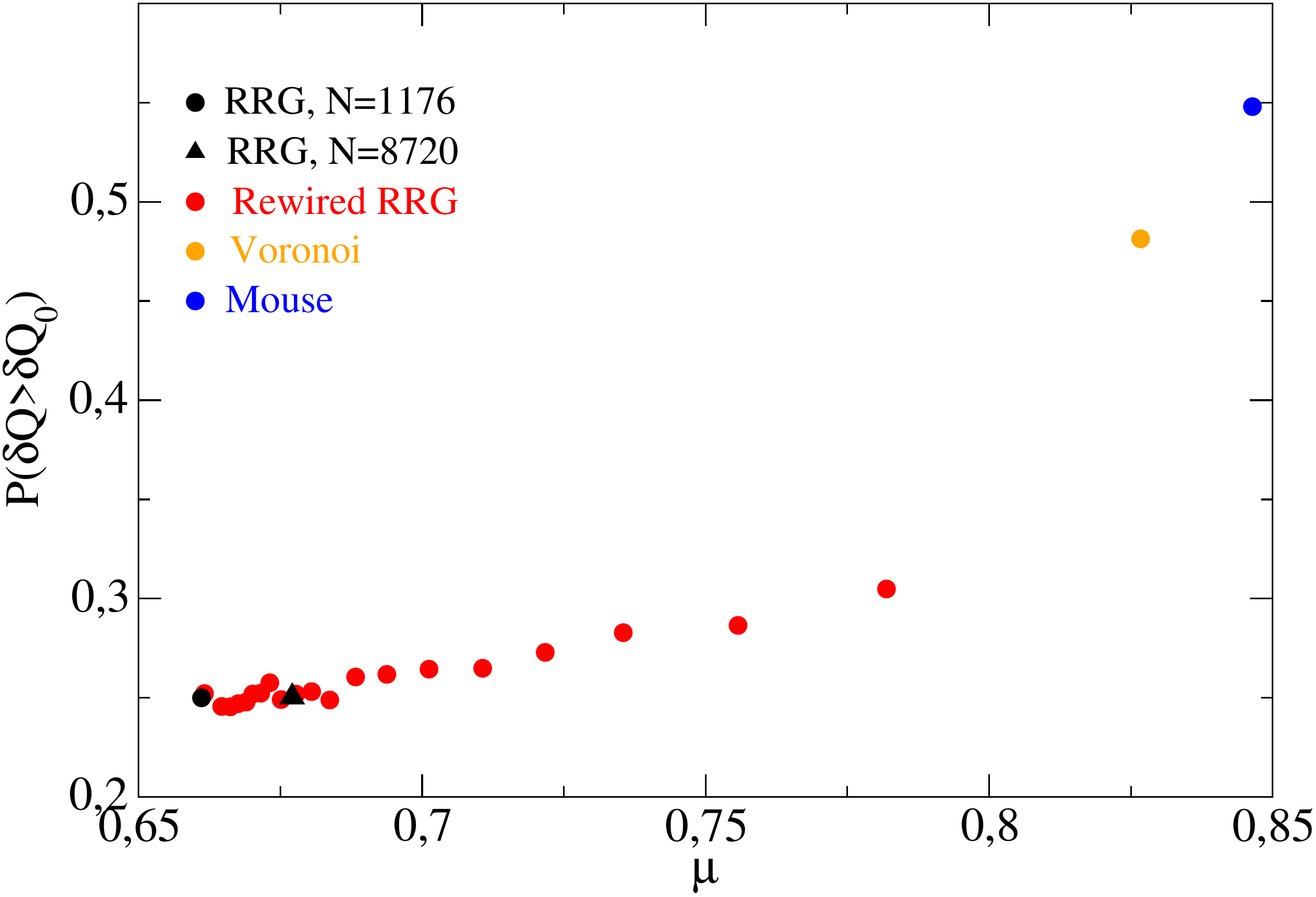}\\
\caption{\textbf{Probability of flow reductions larger than flow threshold $\delta Q_0$ as a function of modularity.} $\delta Q_0$ is chosen to correspond to the third quartile of the distribution obtained for the RRG of same size, as displayed in Fig.~\ref{figdeltaq} by dashed vertical lines.
Same color code as  Fig.~\ref{figdeltaq} (RRG with $N_1=1176$ is the black circle, RRG with $N_2=8720$ is the black triangle. %For the larger graphs, distributions have been shifted according to the same procedure as in Fig.~\ref{figdeltaq} to factor out the size effects. \SLNB{Je propose de supprimer cette dernière phrase, qui me semble inutile vu la définition du seuil de débit, et la note de bas de page ajoutée à ce propos dans le main text.}
\label{figdeltaqbis}}
\end{center}
\end{figure}

%In Fig.~\ref{figdeltaq} we plot the histogram of the $\delta Q_{ij}$.  %There is no general result which describe this distribution if the $q_{ij}$ and $t_{ij}$ are both Cauchy. In Appendix~\ref{removal}, we nevertheless show that for a generic graph, one might expect the  right tail to follow $P(\ln r)\sim 1/r^{1/2}$, and the left tail $P(\ln r)\sim 1/r^{-1/2}$. 
As shown in Fig.~\ref{figdeltaq} (insets), this asymptotic behavior is indeed correctly reproduced for the elementary RGGs (black lines), so that the distribution of $\ln \delta Q$ is approximately symmetric. For these structureless graphs, it is also obvious that the position of its maximum, denoted by $\ln \widehat{\delta Q}$, strongly depends on the network size (compare black lines in top ($N_1$=1176) versus bottom ($N_2$=8720) panel in  Fig.~\ref{figdeltaq}). As shown in Appendix~E for large RRGs, we expect $\widehat{\delta Q}$ to scale as $1/N^2$. Thus, the two distributions obtained for RRGs with sizes $N_1=1176$ and $N_2=8720$ %(black lines in top and bottom panels of Fig.~\ref{figdeltaq}, respectively), 
should match by a translation of the former by $N_2^2/N_1^2$ to the left. This is verified in the bottom panel of Fig.~\ref{figdeltaq} (superimposed black line and black dots) and suggests that, for these structureless graphs, %as a first approximation, 
the distribution of flow reductions is indeed controlled by the distribution of flow rates throughout the network. 
However, whatever the graph size, the distributions of $\ln \delta Q$ become distorted towards larger values when the modularity increases, i.e.~for graphs with substructures. As a result, the corresponding distributions slowly depart from the above predicted scalings, especially for large values of $\delta Q_{ij}$, and the location of their maxima increases. This is consistent with the concomitant increase of $a$, the width of the distribution of the denominator in Eq.~\eqref{qqpgeneral}, demonstrated in the previous Section. 

%The distribution is also clearly dependent on the size of the network. For structureless RRGs, this dependence on the size is perfectly accounted for by a simple shift of the maxima, the distribution of different sizes coinciding after this shift. 

%To be able to compare distributions for different sizes, we have performed this shift in Fig.~\ref{figdeltaq} (bottom). 

%As the modularity increases, the distributions slowly departs from this prediction and becomes distorted to some extent.

%The distribution is also clearly dependent on the size of the network. For structureless RRGs, this dependence on the size is perfectly accounted for by a simple shift of the maxima, the distribution of different sizes coinciding after this shift. To be able to compare distributions for different sizes, we have performed this shift in Fig.~\ref{figdeltaq} (bottom).

%In both top and bottom panels, one sees clearly that the distribution is modified for graphs with structure, and spreads towards larger values of $\delta Q_{ij}$ when modularity increases. 

To further quantify the effect of modularity, %of the presence of communuties in the graph, 
we plot in Fig.~\ref{figdeltaqbis}
the probability of flow reductions larger than a given threshold $\delta Q_0$, chosen to correspond to the third quartile of the distribution obtained for the RRG of same size~\footnote{This definition of $\delta Q_0$ is equivalent as translating all distributions of the above panel of Fig.~\ref{figdeltaq} by $-(8720/1176)^2$, consistent with the size dependence obtained for RRGs in appendix~E.}. From this arbitrary choice, this proportion is obviously 25\% for both RRGs. It increases almost linearly for increasing modularities, up to $\sim$30\% for the rewired RRGs. It even increases by almost twofold for the Voronoi and mouse networks. \SL{Consistent with the results presented in Fig.~\ref{figmodcau}, this larger increase is likely due to topological peculiarities of the community structure in the Voronoi and mouse networks (e.g. heterogeneous distribution of community size, increased probability of inter-links connecting neighbor communities in the 3D space), but additional biological data are needed to explore this point.}

%\OG{A similar effect was seen in the data shown above in Fig.~\ref{figmodcau}. Again, as the Voronoi and mouse networks are just specific instances of the general ensembles of rewired RRGs, this increase can be due to the random dispersion of results in the rewired RRGs, or to a systematic effect due to the peculiarities of the community structure in such vascular networks. We think additional biological data are needed to settle this point.}

\SL{Overall}, in general, a stronger community structure is clearly associated with a higher probability of large flow reductions when removing one edge. As a result, in brain microvascular networks, the removal (or equivalently the occlusion) of one single vessel yields probabilities of large blood flow reductions that may be more than twice the prediction for graphs without substructures. %\SL{Our interpretation of this result is as follows: because vertices belonging to different communities are less connected to each other than vertices within a given community, occlusions affecting inter-community edges, while occurring with a smaller probability, lead to more drastic spatial flow redistributions
%the ratio of inter- versus intra-community edges is much smaller than one, and 
%if the occlusion affects a vessel linking two different communities,
%the flow will be very much affected since such inter-community edges are in a smaller number and the alternative flow paths can be very different. }
\SL{Our interpretation of this result is that the community structure implies that there are subparts of the network less coupled with each other. If the occlusion affects a vessel linking two different communities, the flow will be very much affected since such inter-community edges are in a smaller number and the alternative flow paths can be very different. }

%the integral of the distribution above a certain threshold (corresponding to the value above which 25\% of the probability lies for the generic RRG). The results presented show that the more the community structure is important (as measured by modularity), the larger the occurence probabilities of large values of $\delta Q_{ij}$. 

%Thus higher modularity (i.e.~stronger community structure) is clearly in general associated with a higher probability of large flow reductions when removing one edge. In the case of brain microvascular networks, the removal (or equivalently the occlusion) of one single vessel yields  probabilities of large blood flow reductions that may be more than twice the prediction for graphs without substructures.}

%in brain microvascular networksThis could have important consequences for biological networks: the removal of links can lead to much higher probabilities of large flow reductions in presence of a community structure.

%%%%%%%%%%%%%%%%%%%%%%%%%%%%%%%%%%%%%%%%%%%%%%%%%%%%%%%%%%%%%%%%%%%%%%%%%
%%%%%%%%%%%%%%%%%%%%%%%%%%%%%%%%%%%%%%%%%%%%%%%%%%%%%%%%%%%%%%%%%%%%%%%%%
\section{Conclusion and perspectives}\label{conclusion}
%%%%%%%%%%%%%%%%%%%%%%%%%%%%%%%%%%%%%%%%%%%%%%%%%%%%%%%%%%%%%%%%%%%%%%%%%
%%%%%%%%%%%%%%%%%%%%%%%%%%%%%%%%%%%%%%%%%%%%%%%%%%%%%%%%%%%%%%%%%%%%%%%%%

We have investigated the topological structure of cerebrovascular networks, based on an anatomical network extracted from a mouse brain, model capillary networks derived from space-filling Voronoi tessellations and different graph models belonging to the class of Random Regular Graphs. We have shown that the anatomical network contains substructures corresponding to communities of vessels, which are geographically localized in 3D. The proposed models, including such communities, \SL{have been obtained by rewiring RRGs. They thus belong to graph ensembles describing the universal behavior of regular graphs having a given community strength, and it is possible to tune the number of rewirings so as to obtain an average modularity in line with the modularity of any vascular network.}  Moreover, they provide a reasonable representation of the topology of cerebrovascular networks, as highlighted by their
similar loop distributions. \SL{Noteworthily, whether such a generation scheme could be generalized to describe the topological properties of other vascular networks, especially pathological (e.g.~tumor) ones, remains an interesting open question.}

\SL{Using these simple network models, w}e have then studied how the strength of the\SL{ir} communities affects the distribution of global flow reductions when an edge is removed, or, equivalently, when a single vessel is occluded. For that purpose, we have shown that the resulting flow reduction at network scale can be expressed as a function of the initial flow rate in the occluded %obstructed
 vessel and a topological quantity. Both of them display probability distributions with large tails, resulting in a large probability of rare events.  The presence of community structures even enlarges the tail of the distribution of the topological quantity, resulting  in general in a distribution of flow reductions shifted towards larger and larger values when the community structure becomes more pronounced. In other words, the community structure weakens the network resilience to capillary occlusions, by increasing the probability of larger flow reductions, inducing a large variability of the impact of a single vessel occlusion depending on its location in the network.
 
 The proposed theoretical approach neglects the heterogeneity of vessel conductivities. In other words, by contrast to~\cite{Hudetz1993,Pozrikidis2012,gavrilchenko_resilience_2019,Weber_Single}, we ignored the contribution of vessel morphology, including distributions of diameters and lengths. We also ignored the complex rheology of blood~\cite{Pries_review,Secomb_review,Lorthois_Book_2019}. Our results are nevertheless consistent with recent work modeling the impact of single capillary occlusion in highly resolved numerical simulations accounting for blood microvascular rheology in anatomically realistic microvascular networks~\cite{Weber_Single}, focusing on local flow reorganizations in the vicinity of the occluded vessel. Not surprisingly, the results, obtained for a total of 96 occluded capillaries representing less than 1$\%$ of capillaries, exhibited considerable numerical dispersion, which has been pointed out by the authors as a methodological difficulty. Nevertheless, the median volume of the region with flow reductions above 20$\%$ has been shown to increase by a factor 2.5 between capillaries with low initial flow rate %(0.1 to 4.0 $\mu m^3ms^{-1}$)
 and those with high initial flow rate. % (7 to 25 $\mu m^3ms^{-1}$).} %\SLNB{Moreover, the local topology has been show to strongly modulate this volume, with the highest volume obtained when occluding capillaries with two inflows and two outflows.} 
 This suggests that the present theoretical framework, which enables \SLNB{the use of graph models} focusing on network topology, keeps enough physical ingredients to be relevant.

 While this does not sufficiently reduce the complexity of the problem to yield
 a complete theory relating the community structure to the shape of the distribution of flow reductions (providing e.g full analytical derivations),
 %. However, using simple graph models and focusing on network topology 
 it still enables \SLNB{relevant asymptotic scalings to be deduced} for all quantities controlling this distribution, which considerably helps to interpret the numerical results obtained in the mouse and Voronoi networks. Noteworthy, the last decades have seen tremendous progress of in vivo experimental techniques, including multiphoton microscopy and laser-based techniques that enable \SLNB{microvessels in rodent brains to be selectively occluded}~\cite{shih2012_MP}. This offers large possibilities of data collection, which may be useful to validate our findings. In a complementary way, the present theoretical framework may help enrich data interpretation, e.g. by considering the impact of network communities on \SL{spatial} flow \SL{redistribution~\cite{gavrilchenko_resilience_2019}} %reorganizations
 or by enabling the expected broad distribution of flow reductions to be taken into account in the statistical design of the studies. 
 
Moreover, the present work 
 %while we did not provide a complete theory relating the community structure to the shape of the distribution of flow reductions (providing e.g full analytical derivations), the simple theoretical approach and graph models used in this paper yielded relevant asymptotic scalings that helped to interpret the numerical results obtained in the \SL{mouse and Voronoi} networks. \SLNB{Noteworthy, faible impact des heterogénéités de conductance ?} 
provides a theoretical basis for future studies about the 
 %resilience of  cerebrovascular networks to multiple capillary occlusions~\cite{cruz2019neutrophil}. 
impact of multiple capillary occlusions~\cite{cruz2019neutrophil} on cerebrovascular function. 
As mentioned in the Introduction, %this is a central question to
this may help to understand the interplay between hypoperfusion and amyloid-induced neurodegeneration in the onset and progression of AD ~\cite{Attwell_review2020}.  
We may in particular speculate that the microvascular community structure evolves with disease progression. For example, capillary occlusions at early stages of the disease, i.e.~in healthy networks, may strengthen their community structure. This would increase the probability of larger blood flow reductions induced by further occlusions, providing an additional self-amplificatory mechanism in the positive feedback loop linking hypoperfusion and neurodegeneration in AD~\cite{Attwell_review2020}. In the same way, different network organizations in different brain areas (e.g.~primary versus secondary cortex, subcortical regions, hippocampus), which are being uncovered thanks to whole brain post-mortem vascular network reconstructions in rodents~\cite{Kirst2020}, may be a clue to explain their different vulnerabilities, e.g.~understand why the hippocampus is one of the first damaged brain regions exhibiting cognitive deficits in AD.
 % cognitive deficits
 Long-term vascular remodelling, including capillary rarefaction, in normal or pathological aging~\cite{desposito} may also contribute to modify the community structure of vascular networks in the brain, thus providing an additional mechanism which may explain the considerable overlap between vascular pathology and AD~\cite{Iadecola2010,Iadecola2013}. 
 
 To investigate the above assumptions,
 %speculative
 new datasets finely mapping the whole-brain vascular architecture of rodents in normal aging and at different stages of various brain diseases, including AD, are needed. Moreover, the present framework should be enriched to account for multiple occlusions, e.g.~by introducing a perturbation approach valid in the dilute limit, where the removal of edges can still be considered independent of each other. %interactions between occlusions can be neglected.
 Multiple network inlets and outlets should also be considered. The vascular network within the brain cortex is indeed fed and drained by a large number of penetrating arterioles and ascending venules~\cite{Blinder2013,Shih2015_RF},
 %. While a comprehensive series of in vivo experiments~\cite{Nishimura2006_3MS,Nishimura2007_PA,Nguyen2011,Shih2012_SS,Blinder2013} suggested that compensatory collateral flow from other \SL{network inlets} through the capillary bed is insufficient to prevent dramatic damage induced by the occlusion of a single penetrating arteriole, as reviewed in~\cite{Shih2015_RF}, it may still contribute to enhance the network resilience to capillary occlusions.
which may contribute to enhance the network resilience to capillary occlusions~\SL{\cite{gavrilchenko_resilience_2019}}. By contrast, occlusions of penetrating arterioles induce dramatic damage, as shown by a comprehensive series of in vivo experiments~\cite{Nishimura2006_3MS,Nishimura2007_PA,Nguyen2011,Shih2012_SS,Blinder2013}. This has been interpreted as resulting from insufficient compensatory collateral flow from other network inlets through the capillary bed, as reviewed in~\cite{Shih2015_RF}. Our result suggest that it may alternately be understood as resulting from the hierarchical organization of the network, inlet (and outlet) vessels being those carrying the largest flow rates, thus leading to the largest flow reductions, both at network scale and, by extension, recursively in the neighborhood of the occluded vessel. This phenomenon is likely to be increased if conductance heterogenities are taken into account, such vessels displaying the largest conductances, leading to correlations between high flow and high conductance~\SL{\cite{Al-Kilani2008}}, while such correlations are negligible in the capillary network. \SL{In particular, studying whether these high conductance~/~high flow vessels have any relationship with the community structure would be an interesting extension of this work.} Besides AD, this would open perspectives in the context of ischemic stroke, where neutrophil occlusions of up to 30\% of capillary vessel have been recently discovered, preventing reperfusion after recanalization of the upstream cerebral artery~\cite{ElAmki}. 

\section*{Acknowledgements.}
%************************************************************************************************************************
Research reported in this publication was supported by the European Research Council under ERC grant agreements 615102 (BrainMicroFlow) and 648377 (ReactiveFronts) and by the NIH (awards R21CA214299 and 1RF1NS110054). We gratefully acknowledge P.~Blinder, P.~Tsai and D.~Kleinfeld for sharing anatomical networks. \SL{We also thank the anonymous reviewers for their constructive suggestions of improvements.} OG wishes to thank Laboratoire de Physique Th\'eorique (IRSAMC, Toulouse) for their kind hospitality.

\appendix

%%%%%%%%%%%%%%%%%%%%%%%%%%%%%%%%%%%%%%%%%%%%%%%%%%%%%%%%%%%%%%%%%%%%%%%%%
\section{Details on the construction of network models}
\label{construction}
%%%%%%%%%%%%%%%%%%%%%%%%%%%%%%%%%%%%%%%%%%%%%%%%%%%%%%%%%%%%%%%%%%%%%%%%%
 
%In this section, we review the construction of the networks models used in the present paper. 

All network models considered in the present paper are defined using a graph description of their topology, i.e.~including a 
%Graphs are defined by a 
set of $N$ {\it vertices} connected by {\it edges}. Thus, each edge $i-j$ is uniquely defined by its two endpoint vertices $i$ and $j$. In addition, when the network model is embedded in the 3D physical space, all vertices $i$ have distinct coordinates $(x_i,y_i,z_i)$.

There is a finite number of graphs with a fixed number $N$ of vertices. If we assign a certain probability to each of them, we get ensembles of {\it random graphs}. Random graphs allow to make generic statistical predictions on real-life systems (see~\cite{ReviewGraphs} for a review). The most popular models are regular graphs~\cite{bollobas} (in which each vertex has the same connectivity $z$), Erd\"os-R\'enyi graphs~\cite{ErdRen59,erdos}, or scale-free graphs~\cite{AlbertBarabasi}, depending on the problem under consideration. Here we consider the following models.\\

{\bf Random regular graphs}.  Random regular graphs (RRGs) denotes a subset of graphs uniformly distributed over the finite set of random regular graphs of size $N$ with a given connectivity $z$. Algorithms have been proposed to randomly generate such a subset~\cite{kim2003}. In practice, we use the RandomGraph function of Mathematica.  %such that each vertex has the same connectivity $z$. Algorithms exist that generate graphs uniformly distributed over the set of random regular graphs with a given connectivity~\cite{kim2003}. 
Note that for a graph with $N$ vertices, there are in total $zN$ edge endpoints, and since each edge has two endpoints, $zN$ must be even. In particular, for $z=3$, as is the case in this paper, we must choose $N$ to be even.\\

{\bf Rewired random regular graphs}. We start from a set of $n_c$ independent RRGs of arbitrary even size, \SL{fully disconnected from each other}. We then pick a pair of edges $(i-j,i'-j')$ at random and replace it with $(i-j',i'-j)$.
\SL{This process may connect two initially disconnected components.} We iterate this random rewiring $n_r$ times. At moderate 
$n_r$, the graph keeps some remnant features of its initial elementary components, while for $n_r\to\infty$, it behaves like a single RRG. The initial $n_c$ RRG graphs may have all the same size or different ones. In most numerical applications, we  will consider a rewired RRG built from 20 elementary RRGs of sizes 102, 96, 80, 76, 74, 74, 72, 68, 68, 64, 60, 56, 52, 48, 44, 42, 30, 18, 16, 6. These sizes correspond to those of the communities of one of the Voronoi graphs (denoted $\mathcal{V}$ in the text and defined below), see caption of Fig.~\ref{communautes}, with odd sizes rounded off to even numbers. Such a rewired RRG with $k$ rewirings will be denoted $\mathcal{R}_k$. \\

{\bf Voronoi graphs}. In addition to having mostly 3-connected vertices, the network of capillary vessels, i.e.~the smallest vessels within the brain cortex, is space-filling~\cite{Lorthois_JTB_2010,Smith2019}. This last property can be reproduced by constructing 3D Voronoi diagrams from sets of seed points randomly distributed under the strong constraint that there is only one seed point in each cube of a 3D cubic grid. However, these Voronoi diagrams have high connectivity, with many vertices of degree up to $5$. By randomly merging, pruning or adding vertices following the geometrical constraints described in~\cite{Smith2019}, we get 3D model networks statistically reproducing most of the morphological, topological and transport properties of brain capillary networks~\cite{Smith2019}. Because these networks are generated in a 3D cubic region, all boundary edges are cut and dangling, so that their outer boundary vertex is only 1-connected.
In the present paper, we recursively remove these dangling edges, so that all remaining vertices are at least of degree 2. For simplicity, the resulting graphs are described as \textit{Voronoi graphs} in the present paper. One of them, which we denote $\mathcal{V}$, serves as an illustration throughout the paper.
\\

{\bf Mouse graph}. We use the graph description of a large postmortem dataset ($\sim 1$~mm$^3$ and~$\sim 15,000$ vessel segments) from the mouse vibrissa primary sensory (vS1) cortex previously obtained by~\cite{Tsai2009,Blinder2013}. While this dataset contains vessel diameters and labels classifying vessels in arterioles, capillaries and venules, we discard this information and consider that all edges are equivalent, with unit conductivities. As above, we recursively remove all dangling edges.

%%%%%%%%%%%%%%%%%%%%%%%%%%%%%%%%%%%%%%%%%%%%%%%%%%%%%%%%%%%%%%%%%%%%%%%%%
\section{Communities and modularity}
\label{appcommunities}
%%%%%%%%%%%%%%%%%%%%%%%%%%%%%%%%%%%%%%%%%%%%%%%%%%%%%%%%%%%%%%%%%%%%%%%%%

%\SLNB{Je vous laisse aussi reprendre cette Section avant de l'éditer.}

A graph $\cg$ with $N$ vertices can be partitioned into communities, which are subsets $\calc_k$ of vertices. Intuitively, a community $\calc_k$ is such that vertices in $\calc_k$ are highly connected to one another, with comparatively fewer edges connecting them to vertices outside the community. There is however no unique answer to the question of what a meaningful partition of a graph is. Under reasonable assumptions, it is possible to construct many functions that quantify the relevance of a given partition for community detection~\cite{fortunato2010}.

One such function which has been very much used for that purpose in network theory is the modularity~\cite{NewmanGirvan04}, which for a given partition $\calc$ into subsets $\mathcal{C}_k$ is defined as
\begin{equation}
\mu (\mathcal{C}) =\frac{1}{2m}\sum_k\sum_{i,j\in\calc_k}\left(a_{ij}-\frac{d_id_j}{2m}\right).
\end{equation}
In this expression $a_{ij}$ is equal to 1 if there is an edge connecting vertices $i$ and $j$, and to 0 otherwise, $d_i$ is the number of outgoing edges of $i$, $m$ is the total number of edges, and the sum runs over all subsets $\mathcal{C}_k$ in the partition. The sum $\sum_{i,j\in\mathcal{C}_k}a_{ij}$ gives (twice) the number of edges within the set $\mathcal{C}_k$. The number $d_id_j$ is the number of edges that could connect $i$ and $j$ if they were taken at random. The modularity thus compares the mean probability of having an edge of the graph $\cg$ within a set $\mathcal{C}_k$ to the probability of having such an edge in a graph $\cg'$ where all vertices have the same degree as in $\cg$ but edges are chosen at random.

The various possible functions characterizing the relevance of the partition lead to different methods for community detection in graphs (for a review see~\cite{fortunato2010}). We used modularity-based clustering algorithms, which go through the space of possible partitions trying to maximize the modularity. In practice, we use the FindGraphCommunities function of Mathematica. The algorithm finds the optimal partition with the largest modularity for a given graph. This enables  \SLNB{definition of} the modularity of the graph, which we identify with the modularity of this optimal partition, or equivalently the maximal modularity over all partitions $\mathcal{C}$ :
\begin{equation}
\label{maxmod}
\mu = \mbox{Max}_\mathcal{C}\mu(\calc).
%\frac{1}{2m}\mbox{Max}_\mathcal{C}\sum_{i,j\in\calc}\left(a_{ij}-\frac{d_id_j}{2m}\right).
\end{equation}
We then define the community structure of a graph as the partition with this maximal modularity \eqref{maxmod}.

%%%%%%%%%%%%%%%%%%%%%%%%%%%%%%%%%%%%%%%%%%%%%%%%%%%%%%%%%%%%%%%%%%%%%%%%%
\section{Removing one edge}
\label{removeone}
%%%%%%%%%%%%%%%%%%%%%%%%%%%%%%%%%%%%%%%%%%%%%%%%%%%%%%%%%%%%%%%%%%%%%%%%%
We recall that potentials are solution of $Ap=b$, where
$A$ is the matrix of conductances and $b$ is the given by the right-hand side of Eq.~\eqref{kirchhoff}, namely $b=(0,\ldots,0,-\gamma_{I'}p_{I'},-\gamma_{O'} p_{O'})^T$ (i.e.~potentials $p_{I'}$ and $p_{O'}$ are imposed at inlet vertex $I'$ and outlet vertex $O'$, respectively, see Fig.~\ref{fleches}). We now examine the consequence, on the total flow rate $Q$, of removing one edge from the network. Let us first suppose, without loss of generality, that there is some edge connecting vertices $1$ and $2$ and that we remove it. Let us denote by $A'$ the matrix Eq.~\eqref{mij}  with edge 1-2 removed. The new potentials $p'_j$ in the absence of edge 1-2 are solution of $A'p'=b$.

Matrices $A$ and $A'$ only differ by their upper-left $2\times 2$ corner: the off-diagonal elements are indeed given by $A_{12}=\gamma_{12}$ when edge 1-2 is present, and $A'_{12}=0$  when it has been removed, while the diagonal elements change from $A_{ii}$ to $A'_{ii}=A_{ii}+\gamma_{12}$.  Introducing the column vector $u$ defined by $u^T=\{1,-1,0,...,0\}$, this can be reexpressed as
\begin{equation}
\label{aapunlien}
A'=A+\gamma_{12} u u^T.
\end{equation}
Therefore, \SL{as noted by \cite{Pozrikidis2012,gavrilchenko_resilience_2019}}, $A'$ is a rank-one extension of $A$. The inverse of such a rank-one extension can be obtained from the Sherman-Morrison formula~\cite{encofmath}
\begin{equation}
\label{sherman}
(A+\gamma_{12}uu^T)^{-1}=A^{-1}-\gamma_{12}\frac{A^{-1}u u^T A^{-1}}{1+\gamma_{12}u^TA^{-1}u}.
\end{equation}
Multiplying both members of this equation by vector $b^T$ on the left and $b$ on the right, we get
\begin{equation}
\label{btp}
b^T p'=b^T p-\gamma_{12}\frac{p^Tu u^T p}{1+\gamma_{12}u^TA^{-1}u}\,,
\end{equation}
where $p'=(A+\gamma_{12}uu^T)^{-1}b$ is the solution to the flow equation with edge 1-2 removed. From the definition of $u$, the scalar product of $p$ and $u$ is $p^Tu=p_1-p_2$. As for the scalar product $b^Tp$, we use the fact that $Q=\gamma_{I'}(p_{I'}-p_I)=\gamma_{O'}(p_{O}-p_{O'})=\Gamma\delta P$, which leads to 
\begin{align}
    p_I&=p_{I'}-\frac{\Gamma}{\gamma_{I'}}\delta P\nonumber\\
     p_O&=p_{O'}+\frac{\Gamma}{\gamma_{O'}}\delta P. 
\end{align}
Then, recalling that  $\delta P=p_{I'}-p_{O'}$,
\begin{align}
b^T p&=-\gamma_{I'}p_{I}p_{I'}-\gamma_{O'}p_{O}p_{O'}\\
&=-\gamma_{I'}p_{I'}^2-\gamma_{O'}p_{O'}^2+\Gamma(\delta P)^2.
\end{align}
Equation \eqref{btp} then directly yields
\begin{equation}
\label{qqp}
\delta P (Q-Q'_{(12)})=\frac{\gamma_{12}(p_1-p_2)^2}{1+\gamma_{12}[(A^{-1})_{11}+(A^{-1})_{22}-2(A^{-1})_{12}]}\,,
\end{equation}
where $Q'_{12}$ is the new total flow after removal of edge 1-2.

%\SL{it can be expressed as $D=Q/\Gamma$, where $\Gamma$ is the network equivalent conductance (before edge removal)}.
%The constants $D=p_{I'}-p_{O'}$ and $C=p_{I'}+p_{O'}$ are fixed by the potentials imposed. Since $Q=p_{I'}-p_{I}=p_{O}-p_{O'}$, we have $p_{I}=C-p_{O}$ and $Q=\frac12(D-C)+p_O$. In our notation, $b^T p=-p_{I'}p_{I}-p_{O'}p_{O}$, which can be expressed solely in terms of the constants $C$ and $D$ and the unknown $p_O$. Eliminating the latter, we get the relation
%\begin{equation}
%b^T p=-\frac{C^2+D^2}{2}+D Q.
%\end{equation}
This equation is exact, and since there is nothing special about vertices 1 and 2, it remains valid for any arbitrary edge removed.  %\SLNB{Besides, $\delta P$ corresponds to the total imposed potential difference throughout the network, so 
Thus, in general, we have 
%we can define a network equivalent conductance (before edge removal) as $\Gamma=Q/D$. Finally, for edges with unit conductivities, $q_{ij}=(p_i-p_j)$. 
\begin{equation}
\label{qqp_dim}
\frac{\delta Q_{ij} }{Q} =\frac{\gamma_{ij}}{\Gamma}\frac{1}{t_{ij}}\left(\frac{p_i-p_j}{\delta P}\right)^2\,,
\end{equation}
where  $\delta Q_{ij}=Q-Q'_{ij}$, and $t_{ij}=1+\gamma_{ij}[(A^{-1})_{ii}+(A^{-1})_{jj}-2(A^{-1})_{ij}]$ only depends on the network topology and is  dimensionless. This equation is homogeneous and leads to Eq.~\eqref{qqpgeneral_dim}. 
%Note that in the main text we take $\gamma_{ij}=1$ and $\delta P=1$.

%Since the global potential difference $D=p_{I'}-p_{O'}$ only yields an overall factor, we can set it to 1, hence Eq.~\eqref{qqpgeneral}.

%%%%%%%%%%%%%%%%%%%%%%%%%%%%%%%%%%%%%%%%%%%%%%%%%%%%%%%%%%%%%%%%%%%%%%%%%
\section{Typical value of $t_{ij}$}
\label{meantij}
%%%%%%%%%%%%%%%%%%%%%%%%%%%%%%%%%%%%%%%%%%%%%%%%%%%%%%%%%%%%%%%%%%%%%%%%%
In this section we want to estimate the typical value of denominators $t_{ij}$ in Eq.~\eqref{qqpgeneral}. We will do so by following a reasoning analogous to the ones in~\cite{Kirk73,Pozrikidis2012}. As we shall see, this reasoning is general and does not depend on the values of the edge conductances $\gamma_{ij}$; we therefore keep them in this Appendix.

As in Appendix~\ref{removeone}, we consider, without loss of generality, that the edge is between vertices 1 and 2. The corresponding $t_{ij}$ is defined (see Appendix~\ref{removeone}) as $t_{12}=1+\gamma_{12}[(A^{-1})_{11}+(A^{-1})_{22}-2(A^{-1})_{12}]$. Let $u=(1,-1,0,..,0)$, and denote by $\tilde{p}$ the solution of the equation $A\tilde{p}=\tilde{b}$, where $\tilde{b}$ is a vector such that $p+\tilde{p}=p'$. That is, $\tilde{p}$ is the correction that one has to superimpose to the potential $p$ in order to reproduce the solution $p'$ in the absence of edge 1-2. Since we have
$A(p+\tilde{p})=b+\tilde{b}$ and $A'p'=b$, this leads to $\tilde{b}=(A-A')p'$. Recalling that $A'=A+\gamma_{12}uu^T$, we finally get $\tilde{b}=q u$, with $q=\gamma_{12}(p'_2-p'_1)$.

%$A(p+\tilde{p})=b+\tilde{b}$ and $A'p'=(A+\gamma_{12}uu^T)p'=b$, this leads to 
%$\tilde{b}=q u$ with $q=\gamma_{12}(p'_2-p'_1)$.

The solution $\tilde{p}$ thus corresponds to a pressure distribution where an ingoing edge is attached to vertex $1$ and an outgoing edge to vertex $2$, with an incoming and outgoing flux equal to some value $q_0$ (see Eq.~\eqref{defb}). For $\tilde{p}\ll 1$, we have $q_0\simeq q$.

Using the definition of $t_{ij}$, we then have
\begin{equation}
\label{tildep1p2}
1-t_{12}=-\gamma_{12}u^TA^{-1}u=-\frac{\gamma_{12}}{q}u^T\tilde{p}=-\frac{\gamma_{12}}{q}(\tilde{p}_2-\tilde{p}_1).
\end{equation}
If the network is large and homogeneous enough, the solution $\tilde{p}$ can be seen as obtained from the superposition of a current $q$ coming in at vertex $1$ and going out at infinity, and a current $q$ coming from infinity and going out at vertex $2$. The flux $q$ entering from outside at vertex $1$ will spread evenly along the $z$ wires of the graph connected with it, so that each edge, among which edge $1$-$2$, will carry a flux $q/z$. Similarly the flux outgoing at $2$ will create a flux $q/z$ in all edges arriving at $2$, in particular the edge from $1$ to $2$. By superposition, the flux from $1$ to $2$ will be $\gamma_{12}(\tilde{p}_1-\tilde{p}_2)=2q/z$. Thus, using \eqref{tildep1p2}, we get $t_{ij}=1-2/z$. The prefactor  in Eq.~\eqref{qqpgeneral} is then equal to 
\begin{equation}
\label{qqpgeneralzapp}
\frac{1}{t_{ij}}=\frac{z}{z-2}\,,
\end{equation}
which, for connectivity $z=3$, yields a prefactor 3.
\\

%%%%%%%%%%%%%%%%%%%%%%%%%%%%%%%%%%%%%%%%%%%%%%%%%%%%%%%%%%%%%%%%%%%%%%%%%
\section{Distribution of the logarithm of the flow reduction: asymptotic scaling and position of the maximum}
\label{removal}
%%%%%%%%%%%%%%%%%%%%%%%%%%%%%%%%%%%%%%%%%%%%%%%%%%%%%%%%%%%%%%%%%%%%%%%%%

Equation \eqref{qqpgeneral} relates the flow variation $\delta Q$ due to the removal of an edge with flow rate $q$ as $\delta Q =q^2/t$. As mentioned in Section~\ref{secflux}, both $q$ and $t$ are distributed according to Cauchy distributions and they may be correlated. However, for large RRGs, we have shown that the Cauchy distribution of $t$ is narrow (Fig.~\ref{figcauchy}), so that we can assume %They are a priori correlated; one can however assume, since the Cauchy distribution of $t$ for large graphs is narrow, 
that fluctuations of $t$ are small enough to be neglected compared to flow rate fluctuations. We further assume that the absolute value of $q$ is distributed following %the distribution of $\cp(q)$ of $q$ is 
the Cauchy distribution $\cp(\lvert q \rvert)=\frac{2}{\pi Q_c}\frac{1}{1+(\lvert q \rvert /{Q_c})^2}$
\cite{goirand},
where $Q_c$ is the smallest flow rate characterizing the power law regime. 

Replacing $t$ by its median value, we get from Eq.~\eqref{qqpgeneral}
\begin{eqnarray}
P(\delta Q) = \cp(q)\frac{dq}{d\delta Q}
\sim\frac{1}{\sqrt{\delta Q} \left(1+\delta Q/(3 Q_c^2)\right)}
\end{eqnarray}
for the distribution of $\delta Q$.
Setting $X=\ln(\delta Q)$ and recalling that $p(X)dX=P(\delta Q)d\delta Q$, we have
\begin{eqnarray}
\label{plogdq}
p(X) \sim \frac{e^{X/2}}{1+e^X/(3 Q_c^2)},
\end{eqnarray}
and thus
\begin{eqnarray}
\ln p(X) = a + \frac{X}{2} - \ln\left(1+\frac{e^X}{3Q_c^2}\right),
\end{eqnarray}
where $a$ is some constant that accounts for the prefactor in \eqref{plogdq}. 
At small values of $\delta Q$ we have $X\to -\infty$ and therefore the scaling behaviour
\begin{eqnarray}
\ln p(X) \simeq a + \frac{X}{2}\,.
\end{eqnarray}
At large values of $\delta Q$ we get
\begin{eqnarray}
\ln p(X) \simeq a' - \frac{X}{2},
\end{eqnarray}
with $a'$ some constant.
Besides, the root of the derivative of Eq.~\eqref{plogdq} yields the maximum of the distribution $p$ for $\widehat{\delta Q}=3Q_c^2$. 
Finally, in the limit of large sizes, RRGs of connectivity 3 with one additional inlet and outlet behave like the union of two balanced binary trees of equal height $H$~\cite{encofmath}, with roots corresponding to the inlet and outlet. Thus, we have $N=2(2^H-1)-N_l$, $N_l=2^{(H-1)}$ corresponding to the number of leaves that merge to connect the two trees. For such a graph, the distribution of absolute flow rate is fully described by the power-law regime, so that $Q_c$ is equal to the lowest value of the distribution, i.e.~in the graph leaves. As a result, $Q_c=Q/N_l$. Combining these two  equations leads to $Q_c=3Q/(N+1)$, so that, in the limit of large sizes, $\widehat{\delta Q}$ scales as $1/N^2$.

\end{document}